\documentclass[preprint]{aastex}

\begin{document}
\title{Transmission Spectra as Diagnostics of \\
Extrasolar Giant Planet Atmospheres}
\author{Timothy M. Brown}
\affil{High Altitude Observatory/National Center for Atmospheric 
Research\footnote{The National Center for Atmospheric Research is sponsored
by the National Science Foundation},
3450 Mitchell Lane, Boulder, CO 80307}
\email{timbrown@hao.ucar.edu}



\begin{abstract}
Atmospheres of transiting extrasolar giant planets (EGPs) such
as HD~209458~b must impose features on the spectra of their parent
stars during transits;
these features contain information about the physical conditions
and chemical composition of the atmospheres.
The most convenient observational index showing these features is
the ``spectrum ratio'' $\Re (\lambda)$, defined as the wavelength-dependent ratio
of spectra taken in and out of transit.
The principal source of structure in $\Re$ is the variation with
wavelength of the height at which the EGP atmosphere first becomes opaque
to tangential rays -- one may think of the planet as having different
radii, and hence different transit depths, at each wavelength.
The characteristic depth of absorption lines in $\Re$ scales with the
atmospheric scale height and with the logarithm of the opacity ratio
between continuum and strong lines.
For close-in EGPs, line depths of $10^{-3}$ relative to the stellar
continuum can occur.
The atmospheres of EGPs probably consist mostly of molecular
species, including H$_2$, CO, H$_2$O, and CH$_4$,
while the illuminating flux is characteristic of a Sun-like star.
Thus, the most useful diagnostics are likely to be the near-infrared
bands of these molecules, and the visible/near-IR resonance lines
of the alkali metals.

I describe a model that estimates $\Re(\lambda)$ for EGPs with
prescribed radius, mass, temperature structure,
chemical composition, and cloud properties.
This model assumes hydrostatic and chemical equilibrium in an
atmosphere with chemistry involving only H, C, N, and O.
Other elements (He, Na, K, Si) are included as non-reacting minor
constituents.
Opacity sources include Rayleigh scattering, the strongest lines
of Na and K, collision-induced absorption by H$_2$, 
scattering by cloud particles, and molecular
lines of CO, H$_2$O, and CH$_4$.
The model simulates Doppler shifts from height-dependent winds and
from planetary rotation, and deals in a schematic way with photoionization
of Na and K by the stellar UV flux.

Using this model, I investigated the diagnostic potential of
various spectral features for planets similar to HD~209458~b.
Clouds are the most important determinants of the depth of
features in $\Re$; they decrease the strength of all features as they reach
higher in the atmosphere.
The relative strengths of molecular lines provide diagnostics
for the heavy element abundance, temperature, and the vertical
temperature structure,
although diagnostics for different physical properties tend to be
somewhat degenerate.
Planetary rotation with likely periods leaves a clear signature
on the line profiles, as do winds with speeds comparable
to that of rotation.
Successful use of these diagnostics will require spectral
observations with $S/N$ of $10^3$ or better and resolving
power $R = \lambda/\delta\lambda$ ranging from $10^3$
to $10^6$, depending on the application.
Because of these stringent demands, it will be important to
evolve analysis methods that combine information from many
lines into a few definitive diagnostic indices.
\end{abstract}


\section{Introduction}

The first Jovian-mass substellar object orbiting another star was discovered
about five years ago \citep{may95}.
Since that time, some 50 similar objects have been found, with masses
between about 0.5 and 10 Jupiter masses, and with orbital radii between
.04 and 2.5 AU (e.g., \citet{but97,co97,noy97,que00}).
These objects are commonly termed ``planets'', a terminology to which
I shall adhere, notwithstanding arguments that the objects differ in
important respects from the planets of our own solar system 
\citep{ste00}.
This choice of terminology should not, however, be taken
to minimize the differences
between the extrasolar giant planets (EGPs) and those orbiting the Sun.
Some characteristics of the known EGPs -- notably the tendency of these
massive planets to lie very close to their stars, and the high proportion
of eccentric orbits -- are difficult to understand.
Several plausible theoretical scenarios offer to explain these facts
(e.g. \citet{gol80,lin86,art93,ras96,hol97,maz97,lev98}), 
but distinguishing among them has been difficult because of a
paucity of relevant information.    

Until quite recently, everything that was known about these planets
resulted from precise radial velocity measurements;
assuming the properties of the parent star to be accurately
known, the accessible information was therefore limited to the planet's
minimum mass, along with the semimajor axis, eccentricity, and longitude
of periastron.
Thus, no information about the physical nature (for instance, the chemical
composition) of the radial-velocity EGPs
was available, aside from mass estimates, and these were uncertain
because of unknown orbital inclinations.

This situation changed with the discovery that the companion to the star
HD~209458 transits the disk of the star,
causing periodic dips of about 1.6\% in the flux coming from the star.
Combination of the photometric time series of these transits
with radial velocity data,
in the manner of analysis of other single-line eclipsing binaries,  yields
an accurate estimate of the radius of the planet $r_p$, 
the mass $m_p$, and the orbital inclination $i$
\citep{cha00,hen00,maz00,jha00}.
The estimated values ($r_p$ = 1.55 $\pm$ 0.1 $r_{Jup}$, $m_p$ = 0.69 $\pm$
0.07 $m_{Jup}$, $i$ = 85.9 $\pm$ 0.5 degree)
are consistent with existing models of H/He gas giants that are heavily
irradiated by their parent stars 
\citep{gui96,bur00},
with effective temperatures between 1000 K and 1500 K, depending
upon the size of the planet's orbit, the Bond albedo of its atmosphere,
and the luminosity of its parent star.

Although HD~209458 was the first star to reveal transits by an EGP,
it will almost certainly not be the last.
Given the presence of a planet orbiting a particular star,
and assuming randomly oriented circular orbits,
the probability that transits can be observed is simply
$P_{tran} \simeq {r_* / a}$, where 
$r_*$ is the radius of the star and $a$ is that of the orbit.
As mentioned above, a large fraction of the known EGPs occupy small
orbits, with $0.04~\leq~a~\leq$~0.1~AU.
Most of the parent stars are similar to the Sun in spectral type,
metallicity, and age, and hence have radii that are similar to the Sun's.
In such cases, the probability of a favorable orbital orientation is
about 10\%;
once a few tens of planets had been found, it was likely that one of them
would show transits, and one did.
With new extrasolar planets being discovered at an increasing pace, it
is likely that other transiting planets will soon be found.
Moreover, several searches are underway to detect such planets
directly from their photometric signatures (e.g, the STARE project:
Brown \& Kolinski (1999)
\footnote{Available at 
http://www.hao.ucar.edu/public/research/stare/stare.html}, \citet{bro00};
Vulcan: \citet{bor99}; TEP: \citet{dee98}).
The ultimate extension of such photometric searches will be to
space missions that bypass atmospheric noise sources, and hence will
be able to detect planets as small as Earth circling Sun-like stars
(COROT: \citet{rox99}; MONS: \citet{kje99}; Kepler: \citet{koc98}; 
SPEX: \citet{sch99}).
But for now, the limitations of ground-based photometry (limits not
only on precision but also on duration and continuity of coverage) cause
strong observational biases in favor of objects like that circling
HD~209458.
Thus, in the next few years, most of the transiting extrasolar planets detected 
probably will have approximately Jupiter's radius,
will occupy short-period orbits close to their stars,
and hence will have surface temperatures between about 1000 K and 2000 K.
What can we hope to learn about this class of objects, and what methods
will be useful in studying them?

Transmission spectroscopy is such a method, and one that offers a unique
opportunity to understand the upper layers of EGP atmospheres.
Two circumstances make it an approach worth considering.
First, the high surface temperatures, relatively small masses,
and mostly hydrogen
atmospheres of close-in EGPs imply large atmospheric scale heights.
As much as about 10\% of the planet's cross-sectional area may
be occupied by the outermost (say) 5 scale heights of an EGP
atmosphere, where spectroscopically active species may make an
important difference in the efficiency of light transmission.
The fraction of the star's light that can act as a useful probe of the
planetary atmosphere is thus a few times $10^{-3}$.
This is a small fraction, to be sure, but it is much larger than one would
see in the case of a true Jupiter (which is 10 times colder) transiting
the Sun. 
Second, the measurements that one must perform for transit spectroscopy are
differential in both time and wavelength, 
since they must compare the wavelength dependence of
the spectrum of the star during a transit
with the spectrum seen at other times.
Techniques are available \citep{cha99,col99}
for measuring such time-dependent spectral features with amplitudes of
only a few parts in $10^5$.
For spectral features that span a reasonably large bandwidth, so that
the total photon flux is not too small, one may be fairly confident
of the feasibility of diagnostically powerful measurements.

The purpose of this paper is therefore to investigate the utility of EGP
transmission spectra as diagnostics of the planetary atmosphere's
structure and dynamics.
Since many of the important physical processes in these atmospheres are
poorly understood, my approach is to employ physical constraints
(such as may come from a detailed physical model) where they appear to
be robust, but to parameterize ruthlessly when the behavior of the processes
is uncertain.
In this way I avoid most of the complications inherent in building a
self-consistent model, while retaining some ability to test specific
diagnostics for the many atmospheric features 
that are poorly constrained by theory.
A first step in this direction has been taken by \citet{sea00},
who have conducted a preliminary study of the
features that may appear in EGP transmission spectra.
Their analysis provides a valuable starting point;
most often the conclusions of the current work agree with theirs, though
there are a few important differences.

This paper is organized as follows.
Section 2 gives a qualitative explanation of the phenomena that are
important in deriving the properties of EGP transmission
spectra.
Emphasis is placed on a formalism that allows direct
comparison of model results with a natural observable 
quantity, namely the spectrum ratio $\Re (\lambda)$,
defined as the wavelength-dependent ratio of the flux 
during a transit to that outside it:
$$
\Re (\lambda) \ \equiv \ {F_{transit}(\lambda) \over F_0(\lambda) }\ \ .
\eqno (1)
$$
Practical interest centers on the difference between $\Re$
and unity, which I denote by $\Re^\prime \equiv \Re - 1$.
These ideas are quantified in the form of a numerical model
of $\Re$, which depends upon many parameters of the planetary
atmosphere;
Section 3 describes the parameters defining this model and the
algorithms and data sources used in the computations.  
Section 4 contains the principal results.
First, I describe the features of a fiducial model, whose parameters
correspond so far as can be discerned to those of the companion to HD~209458.
Next I discuss the observable signatures of several variations in the
physical parameters of the model.
Although the sampling in parameter space is far from complete,
the cases chosen allow at least a rough probe of the major physical
variables, and suggest which of these may be estimated reliably, and
which are likely to be ambiguous in their effects.
Section 5 summarizes the previous results, introduces some observational
practicalities, and draws conclusions about promising observational
domains and about the kinds of inferences that may be possible based on
such observations.


\section{Qualitative Description of Phenomena}

The geometry of a transit by a giant planet is illustrated in Figure 1.
For a close-in planet like that circling HD~209458, the orbital period
is on the order of 100 hours, and the orbital radius is about 10 stellar
radii.
The duration of a transit is thus about 3 hours, a time short enough that it
may be observed in its entirety from a single groundbased telescope, but long  
enough to allow low-noise observations.
The planet's speed in such an orbit is typically 150 km s$^{-1}$, assuming the
star to have roughly 1 solar mass;
the variation in planetary Doppler shift during the transit is
then about $\pm$15 km s$^{-1}$, sufficiently large that it must
be allowed for in analyzing observations.
A good observing sequence might be expected to
cover about 7 hours, running from 2 hours before first contact to
an equal time after fourth contact, spanning orbital phases of
$\pm$ 15$^\circ$ around inferior conjunction.
Of course, one could also observe the planet at any
other desired orbital phase, but for purposes of estimating the contribution
to spectral changes from processes other than obstruction by the planetary
atmosphere,
I shall assume the observational coverage just described.
Planets in larger orbits would have transits of longer duration,
scaling as $a^{1/2}$;
for the same temporal duration of observations, the span of orbital phase
would be smaller in proportion to $a^{-3/2}$.

Light comes to the observer from the star/planet system in three
ways, as indicated in Figure 2.
Most of the light is emission from the star that comes directly to the observer.
Smaller amounts are contributed by direct thermal emission from the
planet, and by light from the star that is scattered into the line of sight
by the planet.
Variations in the observed flux with respect to the planet's orbital phase
may come from variations in any of these components.
The spectrum ratio may therefore be written
$$
\Re \ = \ {F_0 \ + \delta F \over F_0} \ = \ 1 \ + \ [ \delta F_{direct} \ + \ 
\delta F_{therm} \ + \ 
\delta F_{scat} ] / F_0 \ \ , \eqno (2)
$$
where the explicit $\lambda$ dependence has been suppressed.
Estimates of the magnitudes of the various terms in Eq. (2)
may be made as follows.

Changes in the direct flux from the star arise from obscuration 
by the intervening planet,
and from changes intrinsic to the star.
Obscuration by the planet is the main topic of this paper.
As mentioned above, for the entire planet the magnitude of the
relative flux change is roughly 1\% to 2\%.
An important contributor to the wavelength dependence of the obscuration
is the center-to-limb dependence of the appearance of stellar
spectrum lines.
This dependence arises from different lines being formed at different depths
in the stellar atmosphere, from anisotropies in the photospheric flows,
and from bulk rotation of the star.
At any instant during a transit,
the planet obscures only a part of the star's visible surface.
The instantaneous spectrum thus represents only the absence of that part
of the star, and not 
an averaged obscuration distributed over the entire disk.
Since line profiles vary across the disk by several tens of percent
(e.g. \citet{mih70}), the instantaneous spectrum may differ from a scaled
version of the unobscured spectrum by several parts in $10^3$.
A time average over the transit reduces these differences but does not
eliminate them, because the average line profiles along the chord 
traversed by the planet need not be the same as the average over the whole
disk area.
Thus, features in the stellar spectrum will tend to appear in modified
form in $\Re(\lambda)$.
To isolate spectrum features caused by the planet's atmosphere, one must either
avoid wavelengths where there are strong stellar lines,
or model the center-to-limb dependence of the stellar spectrum.
Note, however, that distortions of the stellar spectrum may themselves
serve as diagnostics of the stellar atmosphere and of the star's rotation.
The possibilities for such studies deserve careful consideration,
but they are outside the scope of this paper.

A more interesting quantity for the current purposes is 
$(\delta F / F)_{Atmos}$, the part of the obscuration that arises
from rays that pass through the planetary atmosphere.
Its characteristic magnitude is $(\delta A / A)_{Atmos}$,
the cross-sectional area of a
one-scale-height-deep annulus
circling the planetary limb,
divided by the area of the stellar disk:
$$
\left ( {\delta A \over A} \right ) _{Atmos} \ = \ {2 \pi r_p (kT / g \mu) 
\over \pi r_*^2}
\ \ , \eqno (3)
$$
where $T$ is the temperature, $g$ is the surface gravity, and $\mu$ is the
mean molecular weight of the atmospheric constituents.
For $g \ = \ 10^3$ cm s$^{-2}$, $T$ = 1400 K, an atmosphere of
molecular hydrogen, $r_p \ = \ 1.55 \ r_{Jup}$, and $r_* \ = \ 1.3 \ r_{\odot}$,
one obtains a scale height $H$ of 770 km, and for a Sun-sized star,
$(\delta A / A)_{Atmos} \ = \ 2 \times 10^{-4}$.
 
In simple circumstances, one may easily estimate  the difference between
$(\delta F / F)_{Atmos}$ evaluated at two different wavelengths,
using $(\delta A / A)_{Atmos}$ and the corresponding opacity ratios. 
Suppose that $\sigma_1$ and $\sigma_2$ are the opacities per gram of
material at wavelengths 1 and 2,
and further suppose that the densities of the species responsible for the
opacities decrease exponentially with height, with a scale height $H_{\sigma}$.
A tangential optical depth of unity is then reached at different
heights in the two wavelengths, separated by $\delta z \ = \ H_{\sigma}
\ln (\sigma_1 / \sigma_2 )$.
If the opacity sources are uniformly mixed throughout the atmosphere
(which should be nearly the case for the molecules H$_2$O and CO, for instance),
then $H_{\sigma} \approx H$, 
and the difference between the obstructed flux at the two wavelengths
is
$$
\left( {\delta F \over F} \right)_1 \ - \ \left( {\delta F \over F} \right)_2
\ \approx \ \left ( {\delta A \over A } \right )_{Atmos} 
\ln \left ( {\sigma_1 \over \sigma_2} \right ) \ \ . \eqno (4)
$$
In strong atomic or molecular lines, the ratio between the line opacity 
and that of the nearby continuum may easily be $10^4$;
in such a case the height difference between levels of similar tangential 
optical depth is about $10 H$, and the observed line depth
is about $2 \times 10^{-3}$, relative to the total stellar flux.
Since the flux difference depends only logarithmically on the opacity
ratio, extremely large values for the flux difference cannot occur.
On the other hand, weak absorption lines are disproportionately effective,
and they may cause detectable spectrum features,
so long as the lines are strong enough that $\sigma_{line}/
\sigma_{continuum}$ is significantly greater than unity.

For Sun-like stars with moderate levels of magnetic activity, the
intrinsic flux variations should be much like those observed on the Sun.
On timescales of a few hours, these have typical magnitudes of a few
times 10$^{-5}$ \citep{fro97}.
Moreover, these changes largely correspond to small changes in the
solar effective temperature; they have little differential wavelength
dependence.
This source of variation is therefore negligible for the current purposes.

Estimating the change with orbital phase in thermal flux from the planet
requires an assumption about the distribution of temperature across
the planet's surface area.
This in turn depends upon the efficiency of energy transport mechanisms
between the day and night sides, which at present is a matter
of speculation.
For the purposes of an illustrative calculation, I assume that
2/3 of the incident stellar flux is radiated from the planet's day
side, and that the remaining 1/3 is advected to the night side and
radiated from there.
Further assuming uniform temperatures separately on the day and
night hemispheres (i.e., an abrupt temperature jump at the terminator)
results in a day/night temperature difference of about 200 K, given
a mean temperature of 1400 K.
Allowing for the change in projected area of the daylit crescent
as the orbital phase goes from 0$^\circ$ to 15$^\circ$,
the change in thermal emission from the planet would be about 1.3\%.
The change in thermal emission from the planet/star system would be
smaller by a factor of at least 250, because of the ratio of
surface area and temperature of the star to the planet.
Thus, $(\delta F / F)_{therm}$ is at most less than $10^{-4}$
for wavelengths longer than a few microns, and much smaller at shorter
wavelengths.
Because of vertical temperature structure
and strongly wavelength-dependent opacity sources in the planet's
atmosphere, the variable part of the thermal emission will probably
show prominent spectrum lines, just as the transmitted intensity will.
 
An accurate calculation of the variation in scattered light 
with orbital phase would
be fairly elaborate, because of the complicated geometry
of a deep spherical atmosphere illuminated by a star with significant
angular extent \citep{sea00a}.
For an order-of-magnitude estimate, however, consider the difference
between the scattered light from a Lambertian sphere illuminated by a point
source, for orbital phases $\phi$ of 0$^{\circ}$ and 15$^{\circ}$.
At zero phase one observes only the dark side of the planet, so no
scattered flux reaches the observer.
At $\phi \ = \ 15^{\circ}$ one observes a thin illuminated crescent.
The scattered light contribution ${\delta F}_{scat}$
is the product of the incident stellar
flux at the radial distance of the planet from the star, 
the planet's geometric albedo $\alpha$, and the projected area of this crescent.
The scattered light contribution to the relative flux is then
approximately \citep{sob77}
$$
( \delta F / F)_{scat} \ \simeq \ {\alpha \ 
\pi r_p^2 (1 \ - \ \cos \phi ) \over {2 \pi a^2}} \ \ . \eqno (5)
$$
Putting in values appropriate to HD~209458 with $\phi$ = 15$^{\circ}$,
and assuming $\alpha = 1$,
one obtains $(\delta I / I)_{scat} \simeq 10^{-7}$.
If, as theoretical estimates by \citet{sud00} suggest, $\alpha \leq 0.1$
for almost all wavelengths, then the scattered contribution will be
even smaller. 
 
The intensity seen by an observer may also be affected by multiple
scattering processes. 
For instance, starlight may be first reflected
from a deep cloud layer and subsequently directed into the observer's
line of sight by scatterers higher in the atmosphere.
To estimate the importance of this effect, I compare the flux scattered
out of and into a beam directed from the star to 
the observer by a volume element,
where the scattering cross-section within the volume is $\sigma_{scat}$.
The flux removed is then $I_* \sigma_{scat}$, where $I_*$
is the specific intensity in the beam.
Assuming the scattering to be isotropic,
the flux directed into the beam via multiple scattering is $J_{scat} 
\sigma_{scat}$,
where $J_{scat}$ is the mean intensity of already-scattered radiation.
The exact value of $J_{scat}$ depends upon many factors, and could
only be determined by way of the aforementioned elaborate radiative
transfer calculation.
Its maximum possible value, however, is $1/4\pi$ times
the incoming flux from the star (since this is all the light that is
available to be scattered, if none is lost to absorption).
This may be estimated as
$J_{scat} \leq I_* \omega_* /4\pi$,
where $\omega_*$ is the solid angle subtended by the stellar disk,
and $I_*$ is assumed to be a typical stellar intensity value.
For planetary systems such as those considered here, $\omega_*/4\pi \leq 0.01$,
so the multiple scattering contribution is roughly 2 orders of magnitude
smaller than that from single scattering.

I conclude that for wavelengths shorter than the planet's Planck-curve
maximum (roughly 2.5 $\mu$), and excluding the neighborhood of strong
spectral lines in the underlying stellar spectrum, 
$(\delta F / F)_{Atmos}$ is the most important
wavelength-dependent term acting to change the observed intensity
of the star/planet system.
Wavelength variations in this atmospheric 
contribution to the obstructed
flux can almost always be interpreted as the result of opacity-induced
variations in the apparent planetary radius.
Thus, features of a certain depth in $\Re^\prime$ are tied to a
specific level in the atmosphere -- the deeper the feature, the higher
the atmospheric level.
This identification is simple, but important.
It is the only concept required to understand most
of the detailed examples described below.

For the purpose of doing calculations, it is also clear that most of the
machinery of familiar radiative-transfer calculations is unnecessary
at these short wavelengths.
Neither thermal emission nor multiple scattering is important,
and the distinction between absorption and scattering from the beam is likewise
irrelevant;
no important process adds photons to the stellar beam going to the
observer, and all processes that take photons out are equivalent, for
the current purpose.
Thus, one may simply integrate the opacity along each line of sight at
each wavelength, compute the transmissions from the resulting optical
depths, and then perform a suitable average over possible sight lines
to obtain transmission estimates for the planet as a whole.
This is the computational procedure that I shall describe in more detail
below.

At wavelengths longer than 2.5 $\mu$, 
orbital phase variations in thermal emission
may compete with differences due to atmospheric transmission,
depending on (as yet unknown) details of energy distribution and
temperature structure in the planet's atmosphere.
For the purposes of the current study, I shall confine my attention
to wavelengths shortward of 2.5 $\mu$, so that issues of planetary 
thermal emission need not be considered.


\section{Atmospheric Transmission Model}

To assess the utility of diagnostics for EGP atmospheres, 
one needs a quantitative model
of the wavelength-dependent obscuration produced by an EGP silhouetted
in front of its star.
Ideally, such a model would incorporate all of the important physics of
atmospheric structure and radiative transfer in a self-consistent way.
\citet{sea98} and \citet{gou00} have constructed models according to this plan,
and others (e.g. \citet{bur97, bur00a, mar99, sud00}) have explored
much of the relevant physics.
The relevant processes are very complicated, however, and many of them
(such as meteorology and photochemistry) are as yet poorly understood.
Moreover, much of the modeling effort to date has concentrated on the
treatment of isolated brown dwarfs.
These objects are in many ways unsuitable guides to the characteristic
of close-in extrasolar planets, because their atmospheric 
temperature structure is
determined by internal heating rather than by incoming stellar radiation,
and because their large surface gravities lead to relatively small
atmospheric scale heights.
For purposes of estimating the magnitudes of observable effects and
their gross dependences on the nature of the atmosphere,
it is therefore more useful to construct a model that permits easy
parameterization of the atmospheric structure,
and that concentrates on a few of the most important diagnostic
species.
This motivation underlies the model described below,
and is the justification for the restricted set of phenomena considered
in it.

The model calculates the spectroscopic response of the part of the planet's
atmosphere lying near its day/night terminator, since this is
the region that starlight may traverse on its way to the Earth.
In this study I do not consider azimuthal
(ie, latitudinal) variations
in the properties of the atmosphere.


\subsection{Model Input Parameters}

The gross physical characteristics of the planet are determined by
its mass $m_p$, radius $r_p$, and chemical composition.
The mass and radius are described in units of the mass and radius of Jupiter
($M_{Jup}$ = 1.90 $\times 10^{30}$ g = $9.4 \times 10^{-4} \ M_{\odot}$,
$R_{Jup}$= 71500 km = 0.103 $R_{\odot}$).
I define the radius as the distance from the planet's center to a surface
where the pressure is 1 bar,
which happens to be fairly near (but typically below) the level 
at which the tangential optical depth is
unity for the most transparent continuum bands.

The chemical composition of the model planet is defined by the fractional
(by mass) abundances of a few of the most important elements, where
importance derives either from substantial abundance or from the anticipated
spectroscopic activity of the element or its compounds.
For the purposes of this paper I have explicitly allowed for the abundances
of H, He, C, N, O, Na, Si, and K, with relative abundances taken from
\citet{and89}, corresponding to solar abundances.

Simple energy-balance arguments give approximate estimates of the temperature
of giant planets' surface layers.
Nevertheless, the detailed
thermal structure of EGP atmospheres (especially those that are heavily
irradiated by their parent star) cannot yet be calculated with assurance, since
they depend heavily on meteorology, cloud physics, and details of the radiative
transfer.
I therefore simply prescribe the structure in terms of a tabulated relation 
$T(P)$ between temperature and pressure.
I have used for guidance the irradiated 
class IV ``roaster'' model from \citet{sud00},
which is an ad-hoc modification of a non-irradiated model.
Substantial deviations from this reference relation are
possible, as exemplified by existing models that treat radiative
energy balance in a consistent manner \citep{sea98, gou00}.
For instance, the model by \citet{sea98} has temperature decreasing
with decreasing pressure to a greater extent than does that by
Sudarsky et al.,
while that by \citet{gou00} is cooler at its outer boundary,
and its outer layers remain nearly isothermal down to greater depths.
Such differences are characteristic of different treatments of the energy
balance in the planetary atmosphere.

I shall usually describe the temperature structure of model atmospheres
in terms of the temperature $T_0$ at the 1-bar pressure level;
as it happens,
this temperature corresponds fairly closely to the planet's effective
temperature in atmospheres where cloud opacity is not important.
Since the $T(P)$ relation is typically specified on a fairly coarse grid,
I interpolate ln($T$) vs. ln($P$) onto a more convenient grid, with
between 6 and 10 depth points per pressure scale height.
The grid typically extends from a maximum pressure of about 30 bars
(where tangential rays are optically thick at all wavelengths)
to a minimum pressure of $10^{-8}$ bar or less, which is the largest pressure 
at which the most opaque transitions are optically thin.
The pressure grid so defined typically contains 
about 150 distinct depth points. 

The temperature structure is poorly constrained by present knowledge, 
but winds in the atmosphere 
are even more so.
Simple flux redistribution arguments suggest that 
a substantial day/night
energy flow must occur, but, so far as I am aware, no attempts have been
made to calculate the morphology or amplitude of the resulting flow field.
Accordingly, the day/night component of mass flow in the atmosphere is also
given by fiat, in the form of a tabulation giving the line-of-sight
velocity $v(P)$.
Positive $v$ corresponds to a red shift, ie, to a flow from the night side
to the day side of the planet.
  
Clouds are described in terms of parameters that specify the location,
depth, particle size, and particle density
of a predetermined number of cloud layers.
The model expresses
the total mass of particles in a layer in terms of that
determined from the overlying mass of the least abundant (taking into
account stoichiometric proportions) constituent element of the condensate.
For example, in most of the following simulations the uppermost cloud layer
is assumed to consist of enstatite (MgSiO$_3$).  
Since Si is less abundant than Mg in an atmosphere of nominal composition, 
the characteristic cloud mass,
corresponding to a condensation fraction of unity,
is then determined by the mass of Si that lies above the cloud base.
Different cloud masses are permitted: one may specify a 
condensation fraction different from unity,
giving the actual cloud mass relative to the characteristic mass.
One specifies the locations of cloud decks by assigning the pressure level
of the cloud base and the cloud deck thickness measured in pressure scale
heights.
In simple circumstances one expects the cloud base to occur where the
condensate vapor pressure matches its partial pressure, but because of the
substantial variations in this height that may occur because of
weather processes, the model does not insist on this height for the cloud
base.
The opacity is specified by the particle size distribution
and by the substance composing the clouds.
One of three differently-shaped distributions may be chosen 
for each specified cloud deck: 
the ``cloud'',
``log-normal'', and ``haze'' distributions described by \citet{sud00}.
Each of these involves a most-likely particle size, which may be freely specified.

Real clouds are of course far more complicated than those just described.
In practice, however, it turns out that the few quantities 
used in the model are already more
than are really needed for determining the transmission spectrum.
Almost always, the highest layer of cloud is optically very thick 
along a tangential ray at any wavelength,
so that the only parameter of importance is the height of the cloud tops.
Thus, although many details of cloud physics are interesting and important
in other applications (calculating the atmosphere's thermal structure,
albedo, or phase function, for instance), 
these matters are largely irrelevant to the current
discussion.

In addition to specifying properties of the planet, 
its atmosphere, and its star,
one must define the wavenumber (cm$^{-1}$) grid on which the atmospheric 
transmission is to
be computed.
This is done by specifying several wavenumber ranges, each defined by its
minimum and maximum included wavenumber, and by the desired resolution expressed
in velocity units.
Although the calculations were performed on a wavenumber grid, in the plots
that follow, I
display results as a function of wavelength.
The number of wavenumber ranges specified is arbitrary, 
and they may overlap or be
disjoint, as desired.
Within each, the wavenumbers $k$ are assigned with equal separations in ln($k$),
so that constant Doppler resolution is maintained across each range.
This choice simulates the pixel spacing in grating spectrographs, and hence
facilitates comparison of model results with likely observations.
Since the model calculates opacities at discrete wavenumbers (without the
use of opacity distribution functions), one must choose the wavenumber
resolution so that thermally-broadened absorption lines are at least
minimally resolved.
Otherwise, the implicit averaging of opacities across wavenumber bins
leads to significantly incorrect estimates of the emerging flux.


\subsection{Algorithms}

The equilibrium effective temperatures of close-in EGPs are 
expected to be less than about 1500 K;
this is cool enough that their atmospheres should be predominantly
composed of molecular species, rather than atomic ones.
At each pressure level, the model assumes thermodynamic equilibrium
holds among the various molecules formed by the most abundant reactive
atomic constituents: H, C, N, and O, with the
molecules formed then being H$_2$, H$_2$O, CO, CH$_4$, N$_2$, 
and NH$_3$.
I calculate the partial pressures (and hence number densities) 
of these species from the atomic composition and local $T$ and $P$,
using the simple analytic prescription from \citet{bur99}.
This procedure is valid so long as the relative abundances are roughly
solar, in that the number density of O atoms exceeds the sum of
the densities of C and N, and so long as nonequilibrium
and vertical transport processes play minor roles.
Except for certain crude adjustments described below, I do not
account for ionization of atoms or molecules by the incident stellar
UV flux, a process that is likely to be important in the upper atmosphere.

Given $T$ and the composition (hence mean molecular weight) at each pressure
level, and assuming a perfect gas equation of state, 
I compute the
height of each pressure level above the reference (1 bar) level by
integrating the equation of hydrostatic equilibrium.
I allow for the change with height of the gravitational acceleration,
since this can amount to several percent of the total,
but I do not include the height dependence of the planetary mass, since
this variation is much smaller.
The model does not simulate a homeopause, above which different
species show different density scale heights.

I estimate the opacity of the atmospheric material as a function of
wavenumber at each pressure level.
I calculate contributions to the opacity from Rayleigh scattering
by H$_2$ and He,
from resonance scattering by atoms of the alkali metals Na and K,
from absorption by rotation-vibration transitions of the molecules
H$_2$O, CO, and CH$_4$.
and from scattering by cloud particles.
I do not include Raman scattering. 

The Rayleigh scattering opacity $\sigma_R (k)$ (cm$^{-1}$)
is estimated from the atmosphere's local
refractive index $n$, using the expression from \citet{all73},
$$ \sigma_R \ = \ {350 (n-1)^2 \over Lo \ \lambda^4 } \ \ , \eqno (6) $$
where $Lo$ is Loschmidt's number ($Lo = 2.687 \times 10^{19}$ cm$^{-3}$),
and $\lambda$ is given in cm.
I estimated $n$ from the values given in \citet{wea79}, scaled to
the local atmospheric density.
For atmospheres of roughly solar composition, H$_2$ is the dominant
contributor to Rayleigh scattering.
For the planetary models I considered, Rayleigh scattering is important
only for blue and near-UV wavelengths, and fairly deep in the atmospheres.

Collision-induced absorption (CIA) by H$_2$ is an important continuum opacity
source for some regions in the near infrared and for atmospheric pressures
above about 0.1 bar \citep{lin69,bor97}.
Since the opacity from this source scales with the square of the pressure,
it rapidly becomes insignificant at greater heights.
I use Linsky's (1969) parameterizations to estimate the opacity
in the H$_2$ rotation-vibration-translation band around 2.5 $\mu$.
The second overtone rotation-translation-vibration band near 1.2 $\mu$
is about 100 times weaker than the fundamental band, and is not
presently represented in the model.  
 
The alkali metals Na and K should be relatively abundant atomic species in
EGP atmospheres, and they display strong resonance-line
transitions near wavelengths of 332 nm and 589 nm (Na) and 776 nm
and 779 nm (K).
As noted by \citet{bur00,bur00a,sud00,lie00} and others,
at temperatures above 1000 K, these elements do not tend to condense and
precipitate out of the atmosphere, so they provide an important
(perhaps dominant) opacity source through the visible and near-IR
wavelength range.
Strong absorption lines of Na and K are seen in
the emission spectra of the L dwarfs
\citep{rei00}, which have temperatures in the same range as those
expected for close-in EGPs.
I model the opacity using atomic data from \citet{kur85},
usually assuming that all of the atoms are neutral and reside
in their ground states.
These assumptions are probably inaccurate high in the atmosphere, where
vigorous photoionization and low densities (hence low recombination rates)
must lead to significant ionization.
I assume the line shapes of the Na and K lines to be Lorentzian,
with widths given by the quadrature sum of the natural linewidth
(which is almost always unimportant)
and the collisional broadening width.
To estimate the H$_2$-Na collisional cross section, I used the formalism
for computing the Weisskopf radius as described by \citet{bur00a},
giving the collisional broadening as a function of the local density
and temperature.
As explained by \citet{bur00a}, the Lorentzian line shape is only
approximately correct, giving opacity that is too small in the region
just outside the line cores, and too large in the extreme line wings.
The appropriate correction is poorly defined, however, so for the
present the model line profiles retain the simple Lorentzian form. 

As already mentioned,
the current model probably overestimates the
strengths of the alkali metal lines, because atoms of Na and K are easily
ionized by near-ultraviolet radiation from the planet's parent star.
One can make a crude estimate of the effects of this process
by balancing, at each level in the atmosphere, 
the rate of absorption of UV photons
by Na and K atoms against that of radiative recombination with free electrons.
For a few model runs, I have modified the abundances of neutral Na and K
in accordance with this scenario.
For the purposes of this rough calculation,
I made several simplifying assumptions:
(1) I took the only absorbers for stellar photons shortward of the K ionization
edge at 285.6 nm to be K and Na.
This is probably a poor assumption, since the dissociation energies for
some common molecules (notably H$_2$ and H$_2$O) are lower than the
ionization energy for Na.
(2) The ionization of K and Na was taken to be the only 
source of free electrons.
(3) I lumped the stellar UV flux into two wavelength bands, namely that
with $\lambda \leq$ 241.2 nm (able to ionize both Na and K),
and that with 241.2 nm $< \lambda \leq 285.6$ nm (able to ionize only K).
I took the effective photoionization cross-section $\sigma_{bs}$ in each band to
be a flux-weighted average over the band:
$$
\sigma_{bs} \ \equiv \ {1 \over \delta \lambda_b} \int_{\lambda_b}
\sigma_s(\lambda) F_b(\lambda) d\lambda \ \ , \eqno (7)
$$
where $b$ denotes the band and $s$ the species under consideration,
$\delta \lambda_b$ the width of the band, and
$F_b$ the stellar flux in the band, measured outside the atmosphere.
(4) I considered only radiative recombination, excluding other processes
(notably charge-exchange and 3-body processes) by which electrons can
combine with alkali metal ions.
(5) I treated the radiation transfer in a plane-parallel approximation, but with
the external flux incident on the atmosphere at a shallow angle $\theta$.
The angle $\theta$ was chosen so that 
$$
H \sec \theta \ = \ 2 [ (R_p + H)^2 \ - \ R_p^2]
^{1/2} \ \ ,  \eqno (8)
$$
where $R_p$ is the planet's radius and $H$ the pressure scale height,
and where
the right hand side of Eqn (8) is simply the maximum possible 
path length of a ray
lying entirely within a spherical shell of radius $R_p$ and thickness $H$.
(6) I assumed that each UV photon removes only one electron from an atom,
and that recombinations do not create more ionizing photons.
I took photoionization and recombination cross sections from
\citet{lan90,lan91}, and \citet{ver96},
while the UV fluxes were the solar values from \citet{cox00}, suitably
scaled to account for the EGP's smaller orbital distance.
Given the above assumptions and the height dependence of the
Na and K number densities, the computational procedure was to march downward 
from the top of the atmosphere,
at each successive depth solving for the local electron density, 
ion densities, and remaining UV fluxes.

Helium opacity may be quite important
at some wavelengths \citep{sea00}, but it is not included
in the current model.
The reason for excluding it is 
simply that is is comparatively difficult
to calculate the necessary occupation numbers.
Helium is likely to form an extended tenuous exosphere around the planet,
within which photoionization by the stellar flux shortward of 50.3 nm
will be an important process.
Recombination may then populate the high-lying triplet states of the atom,
allowing strong absorption in the 1083 nm line, and possibly also in the
He D3 line at 587.6 nm.
But calculating the populations of the lower levels of these transitions
requires a radiative transfer calculation involving not only He,
but also H$_2$ and H
(which can also absorb these UV photons, and are also very abundant),
and complicated by the near-horizontal incidence of the illuminating
radiation, and by the atmosphere's spherical geometry.
For the purposes of this paper, I therefore simply note that significant
absorption from He is to be expected, and warn the reader
that it does not appear in the model results described below.
 
I estimate molecular opacities for H$_2$O, CO, and CH$_4$,
using line-by-line parameters from the HITRAN 1996 data base \citep{rot98}.
In the wavelength range between 250 nm and 2.5 $\mu$, this involves
approximately 19400 lines of H$_2$O, 400 lines of CO, and 6200 lines
of CH$_4$.
The HITRAN database contains no CH$_4$ lines blueward of 1 $\mu$;
for wavelengths between 300 nm and 1 $\mu$, I have used
low-resolution opacities from \citet{kar94}.
Since the HITRAN database excludes lines with low strengths at its
design temperature of 296 K, there may well be transitions that are
strong at 1400 K but are not represented in the model spectra.
For this reason one should be wary of using the simulated spectra
claculated here in any absolute sense.
For the differential parameter studies that I perform in this paper,
however, a proportion of missing lines is probably not too important.
As with the resonance lines of Na and K, I add each molecular line
into the opacity table with the Doppler-shifted wavelength, strength,
and Lorentzian linewidth appropriate to its pressure level.
Again, I take the line widths to be the quadrature sum of the natural
line width (computed from the tabulated values for transition probability
for each line) and the temperature- and pressure-dependent collisional
linewidths, which are also tabulated for each line.
I use two short-cuts to minimize the computation required for evaluating
the molecular opacity.
First, lines with strengths below a certain threshold (computed separately
at each depth point) are simply ignored.
The threshold is chosen so that
a line with the threshold strength contributes a total optical depth
of $10^{-3}$ in a path length of 1 planetary radius, at the given
density and temperature.
Even with this conservative threshold, in the upper regions of
the atmosphere the vast majority of the tabulated lines disappear from
the calculation, speeding the computation substantially.
Second, lines with widths that are small compared to the wavelength
resolution are approximated as the sum of a central delta function and a
suitably scaled wing, whose total wavelength extent is $\pm~20$ wavelength
bins.
The width cutoff for treating a line in this fashion is chosen so that,
at most, 1\% of the line's opacity is lost in the truncated far wings.
In this approximation, I account for the precise wavelength of line
center by linear interpolation between two similar line profiles separated
by one bin in wavelength.

Cloud particles contribute to the opacity both as scatterers and as absorbers,
with the proportion of absorption depending in detail upon the composition
of the particles.
Fortunately,
for the purpose of calculating the transmission spectrum, 
the distinction between
scattering and absorption is unimportant,
since both remove photons from the beam that continues to the observer.
I therefore approximate the opacity by the scattering component alone,
which I take to be equal to twice the cross-sectional area of the particle
for short wavelengths, going over smoothly to a $\lambda^{-4}$ dependence
for wavelengths that are larger than the particle radius.
I do not include diffraction-related ripples in the scattering opacity
{\it vs.} wavelength relation (such as arise from Mie theory, for instance).
In effect, I assume that the particle-size distribution is broad enough to
smear these features out and lead to a monotonic wavelength dependence.
I compute the ratio of cross-sectional area to mass of particles 
by integrating over the particle-size
distribution, assuming spherical particles with density equal to the bulk
density of the component material.

The opacity contributions just described give the opacity seen in the rest
frame of the corresponding species.
To account for broadening of the line profiles by thermal motions and
by bulk rotation of the planet, I convolve 
the opacity contribution from each species
and at each height with a Gaussian broadening function appropriate
to the temperature at that height and to the molecular weight of the species.
A similar convolution is performed
with a rotational broadening profile 
that is appropriate to a thin annulus rotating about an axis in its own plane.
This profile $\phi _{rot}$ is given by:
$$ \phi_{rot}(dk ) = {1 \over \pi}  \left ( 1 \ - \  {dk \ r_p \over r V_{rot}}
\right) ^{-1/2}\ \ , 
\eqno (9) $$
where $r$ and $r_p$ are the radii of the current pressure level
and of the 1-bar level at which the rotation speed is defined,
and $dk$ is the wavenumber displacement from line center,
measured in velocity units.
Unlike the more familiar stellar rotational profiles, $\phi_{rot}$
is distinctly bimodal, with peaks near $\pm V_{rot}$.
This is a simple consequence of the annular geometry 
of the region traversed by starlight,
which has more area per unit radial velocity near the equator than
near the poles.
Its effect, however, seen at sufficiently high spectral resolution,
is an apparent self-reversal of the line cores (see, e.g., Fig. 16). 
Eq. (9) implicitly assumes that all latitudes have equal opacity at
each depth;  latitude-dependent variations in the composition or
structure of the atmosphere therefore cannot be simulated.

I compute the total opacity $\kappa _{tot} (z,k)$ 
simply by summing the contributions
from the various species (including clouds) at each depth and wavenumber.

Given the opacity vs. depth relation at each wavenumber, I
then compute the tangential optical depth $\tau$
as a function of wavenumber $k$ and of $z^\prime$,
the minimum height of a ray above the 1-bar pressure level:
$$ \tau(z^\prime ,k) \ = \ 2 \int_0^{x_{max}} \kappa (z[z^\prime ,x])
  dx \ \ . \eqno (10) $$
The integral starts with a horizontal ray at the height $z^\prime$,
and continues outward through the spherical atmosphere
until either the total optical depth
exceeds an arbitrary cutoff value of 100, 
or until the ray passes the highest tabulated atmospheric level.
The computation of $z$ at each point along the path
allows for the curvature of the planet's level surfaces and for
the deviation of the initial ray by refraction due to the
atmospere's vertical density gradient \citep{sea00}; 
the geometry of this situation is
illustrated in Figure 4.
At the densities that are of concern
for the transmission problem, the refractive 
deviation is never large enough
to cause significant changes in the computed optical depth,
nor to deflect the ray by a noticeable fraction of the angular diameter
of the illuminating star.
(For example, a ray that is tangent at the 1 bar level is refracted
through an angle of about 3 minutes of arc, as compared to the stellar
angular diameter of more than $12^\circ$.)
Refraction does, however, cause $z^\prime$ to be slightly deeper
(about 2 km, in the case of a ray tangent at the 1 bar level)
than the minimum ray depth $z^*$ that would be projected from the
ray's path outside the atmosphere.
The effective obstructing area of a layer of the atmosphere depends
upon $z^*$, not $z^\prime$, so a secondary result of the integration
in Eq. (10) is to derive the mapping between these two quantities.
For testing, it is sometimes necessary to compute the optical depth
along a vertical ray rather than a tangential one;
a switch allows this to be done when desired. 

As described in section 2,
the quantity one must compute is $\delta F(k) / F(k)$,
the relative diminution of the flux from
the star caused by the presence of the planet.
This is
$$
{\delta F(k) \over F(k) }\ = \ {1 \over \pi r_*^2} \int _0^{z_{max}}
2 \pi (r_p \ + \ z^* ) [1 \ - \ exp(-\tau (z^*,k)] dz^* 
 \ \ . \eqno (11) 
$$
The planet is assumed to be opaque at all wavenumbers for
depths larger than the minimum $z$ treated in the atmospheric model;
the position of the lower boundary must be chosen
deep enough that this is a good approximation.
On the other hand, for computational efficiency it is desirable
to place the lower boundary no deeper than necessary, since
the computing effort required (especially for molecular contributors
to the opacity) grows dramatically with increasing atmospheric
density and pressure.
In practice it is not difficult to guess an appropriate boundary
depth based upon a single trial calculation with a given atmospheric
scale height, or upon the position of the highest opaque cloud
deck.
 
Normal outputs of the model include the run with depth of all of the
thermodynamic and compositional parameters,
tabulations of $\kappa$ against $z$, $k$, and species,
the total opacity $\kappa _{tot} (z,k)$,
the tangential optical depth $\tau (z,k)$,
the mapping $z^*(z)$,
and, finally, the relative obstructed flux
$\delta F(k) / F(k)$. 
 

\section{Results of Model Calculations}


\subsection{Fiducial Model}

As a point of reference, I first constructed a fiducial model intended
to be similar in its important characteristics to the planet of HD~209458.
The parameters of this fiducial model are given in Table 1,
along with those of various models derived from it
(which are described below in more detail).
The model's temperature structure was adapted from the irradiated
Type IV model shown in Figure 1 of \citet{sud00},
the adaptation consisting of raising all temperatures by 200 K in
order to match the desired $T_0$ of 1400 K.
The run of temperature against pressure in the fiducial model
is shown in Figure 5.
Figure 6 shows the run with pressure and height $z$ of the number densities of
important chemical species.
These are mostly uninteresting, with nearly constant mixing ratios 
being the rule.
At the elevated temperatures in the fiducial model, almost all of the
carbon resides in CO, with the remaining oxygen going into H$_2$O;
the number densities of these two molecules turn out to be 
almost equal throughout
the atmosphere.
The total number of these molecules encountered by a light beam that is
tangential at the 1 bar level is quite large:
in terms familiar to IR astronomy, the precipitable water vapor along
a beam reaching to this depth is about 2000 mm.

Methane is the most important exception to the rule of constant mixing ratios.
Its abundance is always small relative to that of CO, but it shows a maximum
around the 1 bar level,
where relatively low $T$ and high $P$ favor its formation.
The total quantity of CH$_4$ is a fairly sensitive function of $T_0$,
increasing as the temperature is lowered below that of the fiducial model.

Clouds in the fiducial model consist of a single
cloud deck of enstatite (MgSiO$_3$),
with a particle size distribution given by the 
\citet{sud00} ``cloud'' distribution,
peaked at 3 $\mu$ size.
The base of this cloud deck lies at $P$ = 0.1 bar, so the cloud tops lie
at about 0.03 bar.
The mass of precipitable material above and contributing to this deck is 
about 0.4 g cm$^{-2}$, great enough that the
total optical depth is very large ($\tau \simeq 500$) at
all wavelengths considered here ($\lambda \leq 2.5$ $\mu$).
The optical behavior of the fiducial model is therefore insensitive
to all features of the atmosphere below the cloud tops.

The fiducial planetary atmosphere rotates with the planet's 3.5-day
orbital period, consistent with tidal locking of the planet's rotation.
The corresponding equatorial rotational speed is
2.0 km s$^{-1}$.
Otherwise, atmospheric flow velocities are taken to be zero at
all levels in the fiducial atmosphere.

Figure 7 shows the transmission spectrum of the fiducial atmosphere binned to
a reciprocal spectral resolution $R = \lambda / \delta \lambda = 3000$.
At the most transparent wavelengths (eg those near 1.0 $\mu$),
the transit intensity is depressed relative to that outside of transit
by approximately 1.53\%;
this is simply the light blocked by the bulk of the planet, which is opaque at
all wavelengths.
For purposes of the following discussion, I will refer to $\Re^\prime$ at
these relatively transparent wavelengths as the ``continuum'',
recognizing that the formation mechanism of these highest parts of
the spectrum ratio is quite different than in the case of self-luminous bodies.

The strongest wavelength-dependent features are the resonance lines of 
neutral Na and K.
These reach depths of 0.2\% of the total stellar flux,
indicating that the planetary radius, seen in the cores of these strong
lines, is some 6\% larger than at transparent wavelengths.
It seems likely that substantial alkali metal features should indeed
be present in the transmission spectrum, but
the very large line depths just described are suspect.
The Na D lines, for instance, are so optically thick that in the model
they remain stubbornly opaque even 
at $P = 10^{-8}$ bar.
This is high enough that photoionization of Na atoms by near-UV flux
from the star is sure to be important;
accordingly, the calculated depths of the alkali metal lines are sure
to be too great.
Moreover, at the relatively low temperatures in the outer layers of the
fiducial model, compounds of Na and K
are stable and may compete for the available alkali
atoms \citep{bur00a}.
Nevertheless, it is plain that spectrum features with
depths of $10^{-3}$ or so are likely to occur.
The strength of these features is similar to
those calculated
by \citet{sea00}.
Perhaps the most important conclusion of the current work is that
relatively large (hence observable) signals are probable, because of
the substantial increase in apparent planetary radius that occurs at wavelengths
where the opacity is high.

Aside from the alkali metals (and helium, which is not modeled here), 
the deepest features in the transmitted spectrum occur in the infrared,
and are caused by H$_2$O, CO, and (to a lesser degree) by CH$_4$.
Most of the features between 0.7 and 2.5 $\mu$
come from the vibration-rotation bands of H$_2$O.
Even in $R = 3000$ spectra, these have obvious signatures.
The band near 1.4 $\mu$, for instance, reaches a depth below the
continuum of 0.08\% to 0.17\% (for $R \ = 3000$ and $R \ = \ 150000$, resp.), 
and has a total equivalent width of about 0.09 nm.
These features will be very difficult to measure from
the ground because of confusion with water vapor in the Earth's
atmosphere.
On the other hand,
it may be possible to isolate lines from high-lying levels that are
not populated at the Earth's low temperature, but become so when $T$
exceeds 1000 K.

A more promising observational target is the first overtone vibration-rotation
band of CO, lying redward of 2.3 $\mu$.
This band is fairly strong and suffers only moderate contamination from
water vapor and perhaps from methane;
as elaborated below, it may provide several useful diagnosics of the atmospheric
structure, especially if observed at high enough resolution to separate
the band into its component lines.
The fundamental vibration-rotation band at 4.8 $\mu$ would, of course, be
even stronger, but it lies in a wavelength range not covered by the current
simulations.
The second overtone band at 1.6 $\mu$ is scarcely visible in the spectrum
of the fiducial atmosphere.

Finally, methane is a minority species in the fiducial atmosphere, particularly
high up where opacity is needed to form strong features in the
transmission spectrum.
Nevertheless, because of its large opacity per molecule,
CH$_4$ absorption is visible in the fiducial atmosphere
at wavelengths longward of 2.2 $\mu$.
The strength of this methane band (and the possible appearance of other,
weaker ones)
turns out to be a sensitive diagnostic of the atmosphere's temperature
\citep{sea99, sea00a}.
Methane has many weaker spectral features at shorter wavelengths, but
because of the low methane concentration, these are not visible in the
fiducial model.

With a cloud particle size of 3 $\mu$, the longest wavelengths considered
here (2.5 $\mu$) are still too short for the cloud deck to become
even partly transparent.
Wavelength-dependent scattering by cloud particles therefore plays no
role in the fiducial spectrum shown in Figure 7.
Similarly, because it scales as the square of the pressure,
and because the cloud tops lie at the relatively low pressure of 0.03 bar,
collision-induced absorption by H$_2$ causes only barely perceptible features in
the fiducial transmission spectrum.


\subsection{Clouds}

Clouds are the most important atmospheric feature determining the nature of
the transmission spectrum.
Even small amounts of suspended particles can be totally opaque at all
optical and near-IR wavelengths, effectively removing from the spectrum
any features that are formed below the cloud tops.
Figure 8 illustrates this fact, showing $\Re^\prime$ for
cases in which the cloud base lies at {1, 0.1, .01, .001} bar
(models N1, fiducial, N2, and N3, respectively).
In all cases, the cloud deck is assumed to consist of MgSiO$_3$,
with a \citet{sud00} ``cloud'' particle size distribution.
For the three lower cloud decks, the characteristic particle size is
3 $\mu$, while for the deck at 0.001 bar, it is 0.3 $\mu$.

As the cloud deck moves higher in the atmosphere, the average amount of
obstructed light increases and localized features in the spectrum disappear,
as the regions where they are formed become submerged.
The remaining profiles of the resonance lines simultaneously grow
narrower, for the same reason.
In the 0.001-bar example shown in the Figure, essentially all molecular
features blueward of 2 $\mu$ are gone, and the only remaining
wavelength-dependent structures are the cores of the resonance lines.
At wavelengths longer than 2 $\mu$, the continuum level rises and
molecular lines reappear, because the wavelength is substantially
larger than the 0.3-$\mu$ cloud particle size, 
and the cloud deck (which already contains little condensed mass, because
of its high altitude) starts to become transparent.

If the cloud tops lie below about 0.3 bar (ie, cloud base at 1 bar),
the height at which the atmosphere becomes opaque is determined by
opacity sources in the gas phase, and the clouds become irrelevant to the
transmission spectrum (though they may still play an important role in
determining the atmosphere's temperature profile).
For such a deep-lying cloud layer, CIA becomes the dominant continuum
opacity source redward of 1.6 $\mu$.
Its effect can be seen in the sloping continuum level between
1.6 and 2.5 $\mu$.
  

\subsection{Metallicity}

Depending upon their histories of formation and evolution,
the atmospheres of EGPs might be enhanced or depleted in
elements heavier than helium, relative to solar composition.
Figure 9 shows $\Re^\prime$ for cases C1 and C2, having respectively 0.5
and 5 times the abundance of heavy elements
(``metals'', in the parlance of stellar evolution theory)
relative to the standard model.
To a first approximation, the effect of these composition
changes is the same as that of changing the height of the
cloud deck, combined with a small change in the assumed
radius of the planet.
This is unsurprising:  the abundances of the most important
opacity sources tend to scale together with changing
metallicity, moving the pressure level for given tangential
optical depth by a constant factor, more or less independent 
of wavelength.

A few anomalies do appear, however.
First, some continuum regions (near 0.43 $\mu$ and
1.05 $\mu$, for instance) actually become brighter with
increasing metallicity.
The affected wavelengths are those where normally minor opacity sources
dominate, the relevant ones being Rayleigh scattering
at 0.43 $\mu$ and the far wings of the alkali metal resonance lines
at 1.05 $\mu$.
Even in these low-opacity regions, the tangential optical depth
reaches unity at pressures
smaller than 1 bar, and hence at heights above the reference
level that defines the nominal planetary radius.
But increasing the metallicity increases the atmosphere's mean 
molecular weight, which in turn decreases the scale height.
Thus, the physical height at which the atmosphere becomes
opaque moves downward, slightly increasing the transmitted light.

A more interesting effect, with possible diagnostic potential,
concerns the methane abundance.
As illustrated in Fig. 10a, the number density of CO relative to
H$_2$ scales in proportion to the total metal abundance;
although not shown, H$_2$O does the same.
The CH$_4$ number density, however, is essentially independent
of the metallicity.
This behavior can be understood by combining Equations A1 and A2 from
\citet{bur99}.
So long as CH$_4$ contains only a small fraction of the carbon,
these equations can be manipulated to show that
$$
B_{CH_4} \ \simeq \ 2 \epsilon \left ( {A_C - \epsilon \over
A_O \ - \ A_C \ - \ \epsilon} \right ) \ \ , \eqno(12)
$$
where $B_{CH_4}$ is the partial pressure of CH$_4$ relative to
that of H$_2$, $A_C$ and $A_O$ are the abundances by number of carbon
and oxygen relative to hydrogen, and
$\epsilon \ = \ P_{H_2}^2 / 2K_1$,
where $P_{H_2}$ is the partial pressure of H$_2$ and
$K_1$ is the temperature-dependent equilibrium constant for the
reaction connecting CO, H$_2$, H$_2$O and CH$_4$.
Thus, so long as $\epsilon$ is small (which occurs at high temperatures
and low pressures), $B_{CH_4}$ does not depend on the absolute abundances
of carbon and oxygen, but only on their abundances relative to each other.

This constancy of the CH$_4$ abundance has an effect on the spectrum, 
as shown in Fig. 10b.
Increasing the metal abundance by a factor of 10 causes a corresponding
increase in the depth of H$_2$O lines, while the strength of
the (very weak) methane lines remains unchanged.
Thus, if CH$_4$ lines are strong enough to be seen, their
depth relative to other molecular lines depends upon $Z$
(but also strongly on the temperature, as will be illustrated next).


\subsection{Temperature}

The effective temperatures and temperature structures of EGPs
are of great interest, because they reflect the importance
of various energy gain and loss mechanisms in the planetary
atmosphere.
At the simplest level, knowledge of the effective temperature should
allow inferences about the total energy absorbed in the
atmosphere, and hence about the planet's visible and 
near-IR albedo.
More subtly, the variation of temperature with height must
be determined by the height dependence of radiation absorption
and emission (by photochemical products, for example),
and by vertical mass transport in the atmosphere.
At present one can do little but speculate about
such processes;
reliable observational diagnostics will be necessary to
gain any real understanding of them.

Fortunately, the atmosphere's temperature influences 
the transmitted spectrum
in several ways.
The most important of these involve changes in the relative
abundances of molecular species,
changes in the atmospheric scale height, 
and changes in the populations of different molecular energy
levels.
To illustrate these effects, I have computed models in which the
temperature at each pressure is raised (model T+) or lowered
(model T-) by 200 K relative to the fiducial model.

As pointed out by \citet{sea00a}, for temperatures near
1200 K and at pressures somewhat less than 1 bar, 
the relative CH$_4$ abundance depends sensitively on temperature,
since it is in this temperature and pressure range that the
dominant carbon-bearing molecule changes from CO to CH$_4$.
Since both CO and CH$_4$ produce many strong lines, the ratio
between them is a useful temperature indicator.
The same sort of transition occurs for N$_2$ and NH$_3$ 
at lower temperatures,
though in this case the diagnostic power is reduced by the
absence of strong N$_2$ lines at useful wavelengths.
This behavior is illustrated in Figure 11.
A practical complication is that in the visible and near IR,
where most of the stellar flux emerges, the CH$_4$ spectrum
is usually intermingled with strong H$_2$O features.
This will make interpretation of the methane spectrum particularly difficult
for ground-based
observations.

Temperature also influences the transmission spectrum by way of
its influence on the atmospheric scale height.
As described in section 2, the relative depth of different
spectrum features typically scales 
as $H \ln (\sigma_1 / \sigma_2 )$,
where $\sigma_1 / \sigma_2$ is the ratio of the opacities
at the two wavelengths, and $H$ is the density
scale height.
Changing $H$ (which scales linearly with temperature) 
therefore changes the strengths of all spectral features,
simply by changing the cross-sectional area of the atmosphere.
In most cases, this effect is the dominant one.
For instance,
for transitions from the same species arising from lower levels
with similar energies, the opacity ratio depends only upon properties
of the atomic or molecular species involved,
but not upon the temperature.
In such cases, the only important variation in line depth arises from $H$.
For a given surface gravity and mean molecular weight,
$T$ is then indicated by the contrast among features within
a given molecular band.
This dependence is illustrated in the top panel of Fig. 12.
Increasing $H$ causes all lines in the band to deepen,
but strong lines deepen more than shallow ones.
As in previous examples, high resolution spectra make the phenomenon obvious
but are not necessary for its observation,
since this kind of comparison may in principle be made between any
two spectrum regions where the same species dominates the opacity.

A more informative way to display such a comparison is to plot
the difference between $\Re^\prime$ computed for two different
models as a function of the fiducial spectrum ratio $\Re_{fid}^\prime$.
The bottom panel of Fig. 12 illustrates this approach
using the same data as in the top panel.
It shows the spectrum ratio difference 
$\delta \Re^\prime \equiv \Re^\prime - \Re_{fid}^\prime$
for the two modified temperature models T+ and T-;
each plotted point corresponds to one wavelength sample from the top panel.
The near-linear dependence of $\delta \Re^\prime$ on $\Re_{fid}$
shows that uniform changes in temperature cause
all to lines strengthen or weaken in accurate proportion to their
strength in the fiducial model.
The slope of the relation between $\delta\Re^\prime$ and $\Re_{fid}$
is proportional to the relative scale height difference between the
models, and the absence of curvature in the relation arises because
the upper part of the fiducial atmosphere is almost isothermal, with
constant $H$.

Because of their robust physical basis and relative independence of
details of chemical composition,
diagnostics related to the scale height are probably the most
effective temperature indicators explored in this study.

The populations of molecular rotational levels depend on
temperature, and therefore so do the relative strengths of
lines arising from them.
These effects are conveniently illustrated near the bandhead
of the CO first overtone vibration-rotation band near 2.3 $\mu$.
Here one can find lines from significantly different rotational
levels in close wavelength proximity, as illustrated in Fig. 13.
In this case, raising the temperature by 200 K increases the
depth of the high-excitation lines relative to the low-excitation
ones by about 15\%.
Many other applications of this principle are possible, provided
that observations with high enough spectral resolution can be
obtained.
Some lower-resolution diagnostics may also exist, exploiting
the tendency of molecular bands to increase in width with increasing
temperature, as higher energy levels become populated.

It is more difficult to measure temperature gradients
in the atmosphere than it is to measure the mean temperature,
but it may nonetheless be practical to derive some
information about the height dependence.
As with temperature gradient measurements in stars, this
may be done by comparing temperature diagnostics from
weak and strong transitions, probing small and large heights
in the atmosphere.
Figure 14 illustrates how such comparisons may be performed.
The left panel shows the run of temperature with depth for three models:
the fiducial model,
model T+,
and model TS, which has an imposed stratosphere in which the temperature
rises by about 35 K per scale height for heights above the .01 bar level.
The right panel shows the spectrum ratio difference 
$\delta \Re^\prime \equiv \Re^\prime - \Re_{fid}^\prime$
plotted against $\Re_{fid}$
for each of the two modified temperature structures,
in the same manner (and using the same wavelength range and resolution)
as in Fig. 12.
For the T+ case in which the temperature at all depths has been raised 
equally, $\delta \Re^\prime$ is nearly a linear function
of $\Re_{fid}^\prime$.
As already
explained, this behavior arises from
the larger scale height in the high-temperature model.
The scatter around a best-fit straight line implies that not all lines
strengthen equally as the temperature is increased.
This difference is mostly the result of differing lower energy levels
among lines, as seen in Fig. 13.
In the case of the stratospheric temperature rise, the relation
between $\delta \Re^\prime$ and $\Re_{fid}^\prime$
is not at all linear.
Rather, it shows a clear break, with weak lines falling near
zero difference, while strong lines are substantially stronger than
in the fiducial model.
This is exactly what one should expect, since 
it is only the deep lines (intensity bins with 
large absolute values of $\Re_{fid}^\prime$) that
sample the upper atmospheric temperature rise.
Of course, when interpreting real spectra, one will not have available
an observed fiducial spectrum against which comparisons can be made;
this role will necessarily be played by synthetic spectra derived from
accurate models.


\subsection{Winds}

The huge incident heat flux on the day side of a close-in EGP is sure
to drive winds that transport energy, momentum, and chemical species
both horizontally and vertically.
The sound speed in the upper layers of hot EGP atmosperes is
typically 3 to 4 km s$^{-1}$;
flows moving at a substantial fraction of this speed are possible.
Information about the nature of the wind systems may therefore be
crucial for understanding even the gross features of the atmospheric 
structure.
Fortunately, the relatively small thermal and pressure broadening
of molecular lines makes them good probes of the atmosphere's
flow structure, unless near-sonic small-scale flows cause additional
line broadening.
The reader should remember, however, that transmission spectra can
yield information only about the region near the planet's terminator.
Other locations, which may be vital for inferring the nature of
large-scale flow patterns, will not be accessible to this kind of observation.

The wind pattern with the simplest observational signature
would be a constant-velocity flow from (say)
the day side to night side, with continuity enforced by an invisible return flow
at great depth.
Such a flow would produce a uniform blue shift in all spectrum lines.
To measure this requires either absolute Doppler measurements of
the planetary lines, or measurements relative to the underlying stellar 
spectrum, combined with a correction for dynamical effects (e.g.,
convective blue shifts) in the stellar atmosphere.
In either case, an accurate ephemeris for the radial velocity of the planet
(such as now exists for HD~209458~b)
will be needed, since the planet's orbital motion will cause Doppler
shifts of $\pm$15 km s$^{-1}$ during the course of a transit.

Another possibility is that uneven heating and the planet's rotation may
combine to drive zonal winds, probably with latitudinal variations in
the wind speed.
Such zonal winds are, of course, common features of the atmospheres of
planets in our own solar system.
The current model does not allow simulation of this geometry,
but the general nature of the effects can easily be anticipated.
Zonal winds would alter the planet's apparent rotation rate, and
(depending upon the latitudinal variation) might also change the
shape of the rotation broadening profile (Eq. 9).
Because the expected rotational speed is small (if HD~209458~b
is tidally locked to its star, the equatorial
rotation speed is only about 2 km~s$^{-1}$),
changes in the rotational line width induced by winds with
moderate Mach number could be important.

Finally, winds may vary strongly in speed or direction as a function of
height in the atmosphere.
The current model is able to handle this situation. 
Figure 15 shows the CO line profiles resulting from a strongly sheared wind
field, with equal and opposite flow velocities of 5 km s$^{-1}$ above and below
the .01 bar level (model V1). 
Such an extreme wind shear is doubtless unrealistic, since the velocity
difference between the layers exceeds the local sound speed;
it does, however, provide a graphic illustration of the 
spectrum formation processes at work.
With the addition of the winds shown in the left panel,
each spectrum line in the fiducial model becomes two overlapping
components -- a strong blue-shifted one corresponding to the winds in the
upper atmospheric region, and a weaker red-shifted one 
corresponding to the lower.
The value of $\Re^\prime$ at which the red-shifted 
component appears is independent
of the line strength, because it is determined by the height 
of the shear layer.
In a strong line, one is really looking at two concentric shells, 
each of which is
opaque at its own Doppler-shifted wavelength.
In a weak line, such as the one at 2.3026 $\mu$, the depth of the
red-shifted component is (as usual) determined by the height at which the line
becomes opaque.
The blue-shifted component, however, results from partial transmission
through the upper shell, which in this line is semi-transparent.
It is worth noting that this is the {\it only} clear example in this
study in which line strength reflects anything except the physical height
of an optical depth surface in the planetary atmosphere.

Figure 16 shows line profiles that result from a more complicated wind
structure with smaller speeds (model V2).
With peak velocities of $\pm$1 km s$^{-1}$, these winds cause more
subtle distortions of the line profiles.
As in Fig. 15, the two wind zones that lie highest in the atmosphere
(with velocities that are positive between 4 and 5 Mm and negative at
greater heights)
cause displacements in the line profiles at values 
of $\Re^\prime$ that correspond
to their vertical locations.
Lower-lying features do not affect the line profiles, since the absorbers
in these lower layers are shadowed by those with similar Doppler shifts
located at greater heights.

The model is not yet able to treat wind fields that vary with latitude
or along each line of sight,
but from the foregoing discussion, the qualitative effects of such velocity
dispersion are evident.
For winds that vary only with latitude,
each element of area around the planet's limb would remove the same amount of
light from the beam, with the wavelength of the absorbed light depending
upon the Doppler shift appropriate to that element.
The effect would be to broaden the line profiles while
conserving the line's equivalent width.
For wind variations along the line of sight, however, each area element
would remove an amount of light depending upon the 
line-of-sight velocity dispersion for that element.
Thus, in this case the line equivalent width would increase.


\subsection{Photoionization of Alkali Metals}

As described in section 3.2, photoionization of Na and K by ultraviolet
light from the parent star may have an important effect on the neutral-atom
number densities, especially
high in the atmosphere.
This is illustrated in Figure 17, which shows the neutral 
Na and K mixing ratios
with photoionization (model PI),
and also the UV fluxes in two bands (band 1: $\lambda \leq 241.2$ nm; band 2:
$241.2 \rm{nm} < \lambda \leq 285.6$ nm), all as functions of height in
the atmosphere.
If photoionization is ignored, the neutral mixing ratios are constant
at the values seen below 1 Mm in Fig. 17.
Figure 18 shows the resulting changes in the Na D doublet.
The principal effect is a reduction in the depth of the line cores,
with little or no change in the far wings of the lines.

Some of the physical assumptions underlying Figures 17 and 18 are questionable,
so the figures should not be taken too seriously in a quantitative sense.
The qualitative behavior is probably correct, however.
As pointed out by Adam \citet{bur00b}, the essential
physics is the same as in a Str\"omgren sphere:
the Na and K atoms are almost completely ionized in the upper part
of the atmosphere, down to a height such that the total ionization/recombination
rate above that height matches the total flux of UV photons.
Once this height is passed, there are no more ionizing photons available,
and the fraction of neutral atoms quickly approaches unity.
The transition between fully ionized and fully neutral occurs over
a small height range,
largely because of the rapid downward increase in density.
Above the transition height, the fraction of neutral atoms falls as a
power of the local density ($\rho^{-2}$, for the case calculated).
The local maximum in the neutral K density near 4 Mm height results from
the sudden increase in electron density caused by the ionization
of Na, which becomes important at this height.
These additional electrons drive up the K$^+$ recombination rate,
increasing the number of neutral K atoms.
In the calculations leading to Figure 17, the most doubtful process
is that causing recombination, for which I assumed radiative recombination
of electrons onto Na$^+$ or K$^+$ ions.
If the recombination rate is in error but the recombination still depends
on a 2-body process (such as charge exchange from negative ions), 
the transition heights will change but the solution will otherwise look
much the same.
If recombination depends upon a multi-body process, then there will be
changes both in the
transition height and in the power-law behavior above that height.
Neither of these changes will qualitatively affect the appearance of
the resonance line profiles in the spectrum ratio.


\section{Discussion}


\subsection{Practicalities}

The examples presented up to this point suggest that useful diagnostics
for cloud height, temperature, and other physical properties
exist in the spectrum ratio of transiting EGP systems.
The examples are somewhat misleading, however, in that no consideration
has yet been given to the spectral resolution and noise level attainable
with likely telescope/spectrograph combinations.
To see a spectrum closely resembling that in Fig. 16, for example,
would require spectra with $R = 10^6$ and $S/N$ per pixel
(relative to the stellar
continuum background) better than $10^5$.
This level of performance will not be achieved soon, if ever.
What can one do with more realistic observational performance goals?

I consider three examples of practical observations of HD~209458:
first, using the HIRES
spectrograph \citep{vog94} at $R = 60000$
on the Keck 10 m telescope at the wavelength of the
Na D lines,
second, using NIRSPEC on the Keck II telescope \citep{mcl98},
in its high-resolution ($R = 25000$) mode
observing in the CO/H$_2$O/CH$_4$ bands near 2.3 $\mu$,
and third, using the proposed Multi-Object Spectrograph on NGST
\footnote{See http://www.ngst.nasa.gov/cgi-bin/pubdownload?Id=496},
operating at $R = 1000$ at wavelengths between 1.69 and 2.46 $\mu$.
I assume that in all cases the dominant noise source is photon
statistics from the star.
Because of the brightness of HD~209458 (m$_{\rm V}$ = 7.65, m$_{\rm K} 
\simeq 6.2$),
read noise and background noise will not be important for any
configuration.
But reaching the photon-statistics noise limit will also require
suppressing other noise sources that are more difficult to manage.
These include the changing effects of telluric lines, drifts in the
point-spread-function
and wavelength zero point of the spectrograph, fringing and calibration
instabilities in the detector, and more.
Bringing these instrumental noise sources to the necessary levels will be
challenging, but experience suggests that it can be done
\citep{cha99,col99}.

With HIRES in typical seeing, $S/N$ for HD~209458 in the continuum 
near 600 nm is about 250 per pixel in a 240 s exposure
\citep{mar00}, where a pixel spans about 2.4 km s$^{-1}$ in velocity space.
For wavelengths near the centers of the Na D lines, this $S/N$ value
will be reduced by up to a factor of 3 because of the smaller flux
in the lines.
The planet's orbital Doppler shift sweeps planetary spectrum features
across the stellar ones during the course of the transit, however,
so for the D lines I assume a typical $S/N$ of 180 per 240 s exposure.
For a transit lasting 3 hours, the accumulated $S/N$ is then about 1250
per pixel.
A similar calculation for the NIRSPEC spectrograph
indicates that at wavelengths near 2.3 $\mu$
the $S/N$ per pixel (1 pixel width $\simeq$ 5 km s$^{-1}$)
accumulated over a transit should be about 2000.

Figure 19 shows the region around the Na D lines at the HIRES sampling and
$S/N$ estimated above.
This plot assumes the fiducial model with photoionization included (model PI).
Although the noise at any individual pixel is comparable in magnitude
to the total depth of the line profiles, the sampling is sufficiently
fine compared to the profile widths that the lines are easily seen,
even in unsmoothed data.
After smoothing with a 9-pixel triangular filter (giving effective
spectral resolution of $R = 20000$),
the lines are obvious, and after correcting appropriately for the
effects of the smoothing, one can make reasonably accurate
estimates of their depths and widths.

A first attempt to detect line signatures in this fashion
has been made by \citet{bun00}.
They used high-resolution spectra of HD~209458 and of 51 Peg that
were taken in the course of radial velocity programs, and searched
for differences between the spectra obtained during transits
(for HD~209458) or near inferior conjunction (for 51 Peg),
and those obtained at other times.
They detected no signature of the planetary atmospheres,
but because of the relatively small number of spectra available
(only 2 during transit, in the case of HD~209458),
their sensitivity in $\Re ^\prime$ was at best a few percent.
As Fig. 19 illustrates, this sensivitity falls short of what is
needed by a factor of 10 or so.
These results are therefore illuminating as regards technique,
but are not yet precise enough to speak to the nature of the planetary
atmospheres.

The situation is yet more difficult with respect to molecular bands,
because the lines are not so strong.
Figure 20 illustrates a model spectrum between 2.2 and 2.5 $\mu$
at NIRSPEC's resolution,
both without (top) and with (bottom) simulated photon noise.
In this case the model spectrum is that from model T-.
Clearly this example is more problematic, in that individual lines
cannot be distinguished, and even the CO bandhead at 2.3 $\mu$ is
difficult to identify.
Smoothing these data helps in locating the bandhead,
but one can do better by using {\it a priori} knowledge
about the molecular line wavelengths and strengths.
The easiest way to do this is to cross-correlate an observed
spectrum with a template containing expected lines from some
molecular species of interest,
in the fashion described by \citet{wie00}.
Figure 21 shows the results of such cross-correlations between 
templates for the molecules [CO, H$_2$O, CH$_4$] and the simulated noisy
spectra computed for two different atmospheres.
For molecular template spectra $M_s(\lambda)$ for the species $s$, I used 
$$
M_s(\lambda ) = \ln ( \sigma_s (\lambda ) + \sigma_{min} ) \ \ , \eqno(13)
$$
where $\sigma_s(\lambda )$ is the computed opacity for species $s$ at
the fiducial atmosphere's 0.01-bar level, 
and $\sigma_{min}$ is an opacity floor,
chosen a factor of 10 smaller
than the largest continuum opacity in the chosen wavelength band.
Thus, the templates are similar (but not identical) to the spectra
that would result from the model using only continuum opacity sources
and opacity from the molecule under consideration.
The atmosphere models used were the fiducial model and the T-
model whose spectrum is illustrated in Fig. 20.
As discussed in section 4.4, the main differences between the T-
model and the fiducial models are that, in the cooler model, the bands of
CO and H$_2$O are weaker, while CH$_4$ (although still weak compared
to CO and H$_2$O) is greatly strengthened.
This behavior is clearly revealed in the cross-correlation plots of Fig. 21,
in spite of the very substantial photon noise.
The presence in Fig. 21 of obvious correlation peaks at zero lag ({\it i.e.,} at
their expected locations) shows robust identifications of CO and H$_2$O
in the spectra of both atmospheres.
Their correlation peak sizes indicate typical line strengths;
they are larger for the fiducial model than for the T- model, 
as expected.
The correlation peak for CH$_4$, on the other hand, is marginally
detectable for the T- model and absent in the fiducial.
Clearly, interpreting correlation
strengths in a quantitative sense will depend upon comparison with
models that are accurate and complete.
The position and width of the correlation peaks, in contrast, permit
comparatively model-independent inferences about 
the Doppler shift and broadening of the
corresponding line systems.
Displacement of the correlation peak from zero lag implies a systematic
Doppler shift of observed lines relative to the template, and hence
day/night winds.
The precision with which such winds could be measured depends upon the
width of the cross-correlation peak and upon the noise level in the
cross-correlation.
For the CO and H$_2$O examples given, it is about $\pm$0.35 km s$^{-1}$.
The shape of the peak contains information about the intrinsic line
shape, so long as the true line widths are larger than the
instrumental width;
for the examples shown in Fig. 21, this is not the case,
so no line-shape information can be obtained.

Observing transit spectra from space would be highly desirable, because
the principal near-infrared transitions of interest in the planetary
spectrum are contaminated (or obliterated) by absorption in the Earth's
atmosphere.
Unfortunately, such observations cannot be obtained at present.
The Space Telescope Imaging Spectrograph (STIS) on HST does not have
infrared capability, and even in visible light its spectral resolution
is marginal for separating the planet's alkali metal lines from their stellar
counterparts.
The Next Generation Space Telescope (NGST) offers better prospects:
it will be optimized for infrared observations, and its orbit will
allow for uninterrupted observations of entire transits.
Unfortunately, ideal transit observations are characterized by single
targets, high spectral resolution, short exposure times, and very
high $S/N$ ratio, while the NGST's principal scientific goals demand
none of these things.
Thus, instruments on the NGST will probably be poorly optimized for
measuring transit spectra.
It is nonetheless worthwhile to consider what might be observable 
with NGST's planned instruments.
For this purpose, I have concentrated on spectra one might obtain
with the Multi-Object Spectrograph operating at $R = 1000$ (2 pixel) over the
1.69 $\mu$ to 2.46 $\mu$ wavelength range.
With this spectral resolution, the NGST's 8 m aperture, 
and an assumed optical efficiency of 40\%, the greatest
difficulty in observing HD~209458 would be the very short integration 
times allowed before saturating the detector
(roughly 1 s, assuming a detector well depth of $10^5$ e$^-$).
Depending upon the available detector readout options, these short
exposures may lead to very low duty cycles.
Figure 22 shows the $\Re^\prime$ spectra that might be obtained
during a single 3-hour transit of HD~209458~b, under various assumptions
about the duty cycle,
and using the Fiducial model for the planetary atmosphere.
For all of the panels in Fig. 22, I assumed individual integrations
with $S/N = 320$, with integration/readout cycle times for the top three panels
equal to 300 s, 20 s, and 2 s.
As shown in the figure, the corresponding integrated $S/N$ values
are 1900, 7350, and 23250;
the last of these corresponds, within a factor of 1.4, to shot-noise-limited
performance for the 8 m telescope.
It is clear that the presence of molecular bands can be verified and
their strengths estimated even in the noisiest case.
With improved $S/N$ it becomes possible to identify finer details,
until, near the maximum possible efficiency, essentially all of the
detail present in the spectrum at this resolution can be seen.
Not all of the diagnostics described above are feasible at this resolution,
but one could nevertheless learn much from observations of this quality.


\subsection{Prospects}

The present study suggests that spectra of transiting EGPs can be used
to infer many properties of their atmospheres,
including cloud heights, heavy-element abundances, temperature and
vertical temperature stratification, and wind velocities.
The spectrum diagnostics resulting from these physical changes are
to some degree degenerate.
For instance, lower cloud height, higher temperature, higher heavy
element abundance, and greater turbulent velocities all tend to make
spectrum lines stronger.
Differences among the behavior from these various causes do exist,
however, and may be sufficient for their effects to be untangled through
the analysis of data from existing instruments.
At this point the most important uncertainties about the
practicality of this program come from the EGPs
themselves.
Most crucially, are the upper atmospheres of EGPs sufficiently
free of clouds that atomic and molecular lines (upon which most
of the diagnostics depend) can be observed?
Theoretical arguments exist both for the presence and absence of
high-level clouds, but in all likelihood, the only way to answer
such questions is through observations.
Many kinds of observations of HD~209458~b are already being undertaken,
so there should soon be available a better factual basis for
models such as the one described here.
When that basis exists, it will be easier to know which avenues to
follow in improving the models and in developing better observing
and interpretive tools.

\acknowledgements

I am grateful to S. Seager, D. Sasselov, and A. Burrows for
helpful discussions, and to D. Charbonneau and K. MacGregor for their 
careful readings of the paper,
and for many suggestions.
I thank D. Kolinski for assistance in preparing Figures 1, 2, and 3.
This work was supported in part by NASA grant W-19560.



\clearpage

\figcaption{Geometry of a transit by an extrasolar giant planet
(EGP). 
The planet's orbital phase is $\phi$.
(1) shows the situation before the transit, (2) at the center of the
transit.
The scale in these figures and the curvature of the planet's orbit
are distorted for clarity.
\label{fig1}}

\figcaption{
3 sources of radiation come to an
observer from the star/planet system.
Direct radiation from the star 
(D) is partially blocked during the transit.  Thermal radiation
from the planet (T) varies because of presumed temperature
and emissivity variations across the planet's surface.  The
scattered component (S) varies because of the changing illuminated
phase, and because of possible scattering-angle dependences
caused by particles in the planetary atmosphere.  Note that,
because of the angular extent of the parent star,
significantly more than a hemisphere of the planet is
partially illuminated.
\label{fig2}}

\figcaption{Atmospheric opacity (shown as a function of wavelength
in the upper trace) corresponds directly to apparent radius of
the planet, since in opaque regions of the spectrum, one must
go to higher altitude and lower density before tangential rays
are transmitted.
High-opacity regions appear as absorption lines in the spectrum
ratio $\Re^\prime$ because the planet appears larger at these
wavelengths, and blocks more of the star's light.
\label{fig3}}

\figcaption{Geometry of refraction of a ray within the planetary
atmosphere, with the refraction angle greatly exaggerated for clarity.
The radial distance $z$ is the height of the ray above the nominal
surface of the planet (whose radius is $r_p$) 
at an arbitrary point along the ray path.
Similarly, $z^\prime$ is the minimum height actually reached by
the ray, and $z^*$ is the minimum height that would have been reached
in the absence of a refracting atmosphere.
\label{fig4}}

\figcaption{Temperature and pressure profiles with height in the
fiducial model, which has $T_0 \simeq 1400$ K.
The hatched band indicates the height range spanned by
a layer of clouds, consisting of 3 $\mu$ particles of MgSiO$_3$.
\label{fig5}}

\figcaption{Number densities of important atmospheric constituents
{\it vs.} height in the fiducial model.
This model assumes no ionization of any species.
The near-equality between CO and H$_2$O number densities is
coincidental.
Note that, at the temperatures and pressures in this model,
 very little of the carbon resides in CH$_4$, and very
little nitrogen in NH$_3$.
\label{fig6}}
 
\figcaption{Fiducial model spectrum ratio $\Re^\prime$ {\it vs.}
wavelength.
The maximum value of -1.53\% corresponds to the fraction of the
star's light blocked at the wavelengths where the atmospheric opacity
is lowest.
Over most of the wavelength range, the spectrum was computed at 
$R = 150000$, but it is displayed here averaged and rebinned
to $R = 3000$.
Significant atomic and molecular absorption features are labeled.
\label{fig7}}

\figcaption{Dependence of the model $\Re^\prime$ on cloud height, with
cloud base pressures (in bars) labeled.
Cloud tops are 1 pressure scale height above the cloud bases. 
The spectrum does not change further for cloud bases below 1 bar.
The 0.001-bar case assumes 0.3 $\mu$ particle diameters; 
all of the other cases assume 3 $\mu$ particles.
\label{fig8}}

\figcaption{Dependence of $\Re^\prime$ on the abundance of elements
heavier than He.
The thick line (model C1) shows metals reduced by a factor of 2 relative to the
fiducial model;  the thin line (model C2) shows an increase by a factor of 5.
Note expanded scales in both $\lambda$ and in $\Re^\prime$.
\label{fig9}}

\figcaption{Left panel: mixing ratios of CO and CH$_4$ {\it vs.}
height for models C1 (thick lines) and 
C2 (thin lines).
The CO mixing ratio scales with metallicity, while the CH$_4$
mixing ratio changes little.
Right panel: detail of $\Re^\prime$ for low (dot-dashed line)
and high (solid line) metallicity, showing both CH$_4$ and H$_2$O features.
\label{fig10}}

\figcaption{Dependence of H$_2$O line strengths on temperature.
Top: $\Re^\prime$ for the fiducial model (thick solid line) and
for models with temperatures increased (thin solid) and decreased
(dot-dashed) by 200 K (models T+ and T-, respectively).
Bottom: Spectrum ratio difference 
$\delta \Re^\prime = \Re^\prime - \Re^\prime_{fid}$
{\it vs.} $\Re^\prime_{fid}$ for the T- (+ symbols) and T+
(diamonds) models.
The near-linear relation between $\delta \Re^\prime$ and $\Re^\prime_{fid}$
is a signature of differing atmospheric scale heights among the models.
\label{fig11}}

\figcaption{Temperature dependence of CH$_4$ lines.
Plotted values are the difference between $\Re^\prime$ and its
local (in wavelength) maximum for the fiducial model (thick solid line),
and for models T+ (thin solid) and
T- (dot-dashed) by 200 K.
\label{fig12}}

\figcaption{Temperature dependence of CO lines.
Shown are $\Re^\prime$ for the fiducial model (thick line) and models T+
(thin solid line) and T- (dot-dashed line).
All curves have been shifted to give maximum values of zero, and the
T$\pm$ models have also been scaled so that the line depths at
2.3252 $\mu$ and 2.3268 $\mu$ agree with those in the fiducial model.
These adjustments compensate for scale-height influences on the line depths,
which are nearly the same for all lines.
The lower-state energies (in cm$^{-1}$) for the 4 
strongest lines are shown on the Figure.
The weak lines present only in the T- model are from CH$_4$.
\label{fig13}}

\figcaption{Left: Run of temperature with depth for fiducial model (solid line),
model T+ (diamonds), and model
with ``stratospheric'' temperature rise above $P = 0.01$ bar 
(model TS; + symbols).
Right: Spectrum ratio differences $\delta\Re^\prime$ {\it vs.} fiducial
spectrum ratio $\Re^\prime_{fid}$.
Each point represents one wavelength sample with 20 km s$^{-1}$ width
drawn from the H$_2$O band with $1.3 \ \mu < \lambda < 1.4 \ \mu$.
\label{fig14}}

\figcaption{Velocity dependence of CO line shapes.
Left: imposed variation of wind speed with height (model V1).
Right: line shapes and positions for fiducial model (dashed line) and
for model with winds (solid line).
\label{fig15}}

\figcaption{As Figure 13, but with more complicated and smaller-amplitude
imposed wind field (model V2).
\label{fig16}}

\figcaption{Height dependence of calculated neutral atom mixing ratios
and UV fluxes, allowing for photoionization and radiative recombination
of Na and K (model PI).
The wavelength ranges spanned by the UV bands are described in the text.
\label{fig17}}

\figcaption{Effect of photoionization as shown in Fig. 15 on the Na D line
profiles.
Curves show $\Re^\prime$ with (solid lines) and without (dashed lines)
photoionization.
The inner box shows a magnified view of the region near the line cores.
\label{fig18}}

\figcaption{Spectrum ratio $\Re^\prime$ in the region of the 
Na D lines for model PI, as it might be
observed with the HIRES spectrograph.
Resolution is 2 km s$^{-1}$ per sample, and simulated photon noise
(independent Gaussian-distributed with rms=0.08\%) has been added.
Top: dots show individual sample values, solid line shows the result of
smoothing with a 9-sample-wide triangle filter.
Bottom: magnified view of line core region, with the addition of the
noise-free model profile (thick solid line).
\label{fig19}}

\figcaption[figs/fig19.eps]
{Spectrum ratio $\Re^\prime$ in the region of the first
overtone CO bandhead, as it might be seen with the NIRSPEC spectrograph
in high-resolution mode ($R = 25000$, 12 km s$^{-1}$ per resolution element).
Top: noise-free spectrum ratio for model T-.
Bottom: same as top, with addition of simulated photon noise with
rms=0.033\%.
\label{fig20}}

\figcaption[figs/fig20.eps]
{Cross-correlation of spectrum templates for CO, H$_2$O,
and CH$_4$ with noisy $\Re^\prime$ spectra for the fiducial model (left
column) and for model T- (right
column).
Note change of vertical scale for the CH$_4$ plots.
\label{fig21}}

\figcaption[figs/fig21.eps]
{A portion of the infrared $\Re^\prime$ spectrum of the fiducial model,
as it might be observed during a single transit with the Multi-Object
Spectrograph on the NGST at $R = 1000$.
The top 3
panels show different assumed noise levels, resulting from
different assumed integration/readout cycle times: from the top,
300 s, 20 s, and 1 s.
The bottom panel shows the noise-free spectrum.}
\clearpage

\begin{deluxetable}{lcccccccccccc}
\tabletypesize{\small}
\tablecaption{Input Parameters for Models. \label{tbl-1}}
\tablewidth{0pt}
\tablehead{
\colhead{ } & \colhead{Fid} & \colhead{N1} & \colhead{N2} &
\colhead{N3} & \colhead{C1} & \colhead{C2} & \colhead {T+} &
\colhead{T-} & \colhead{TS} & \colhead{V1} & \colhead{V2} & \colhead{PI}}
\startdata
$R_* \ (R_\odot)$ & 1.54 & & & & & & & & & & & \\
$T_*$ (K) & 6000 & & & & & & & & & & & \\
$a$ (AU) & 0.04 & & & & & & & & & & & \\
$R_p \ (R_{Jup}$) & 1.54 & & & & & & & & & & & \\
$M_p \ (M_{Jup}$) & 0.69 & & & & & & & & & & & \\
$T_0$ (K) & 1400 & & & & & & 1600 & 1200 & & & & \\
$T(P)$ & Std & & & & & & & &Strat& & & \\
$V_{eq}$ (km/s) & 2 & & & & & & & & & & & \\
Cld Top (bar) & .037 & 0.37 & .0037 & .00037 & & & & & & & & \\
$Z/Z_\odot$ & 1.0 & & & & 0.5 & 5. & & & & & & \\
Wind (km/s) & 0 & & & & & & & & & $\pm 5$ & $\pm 1$ & \\
Photoion? & No & & & & & & & & & & & Yes \\
\enddata
\tablecomments{All parameters take their values in the fiducial ``Fid''
column unless noted otherwise.}
\end{deluxetable}

\begin{figure}
\plotone{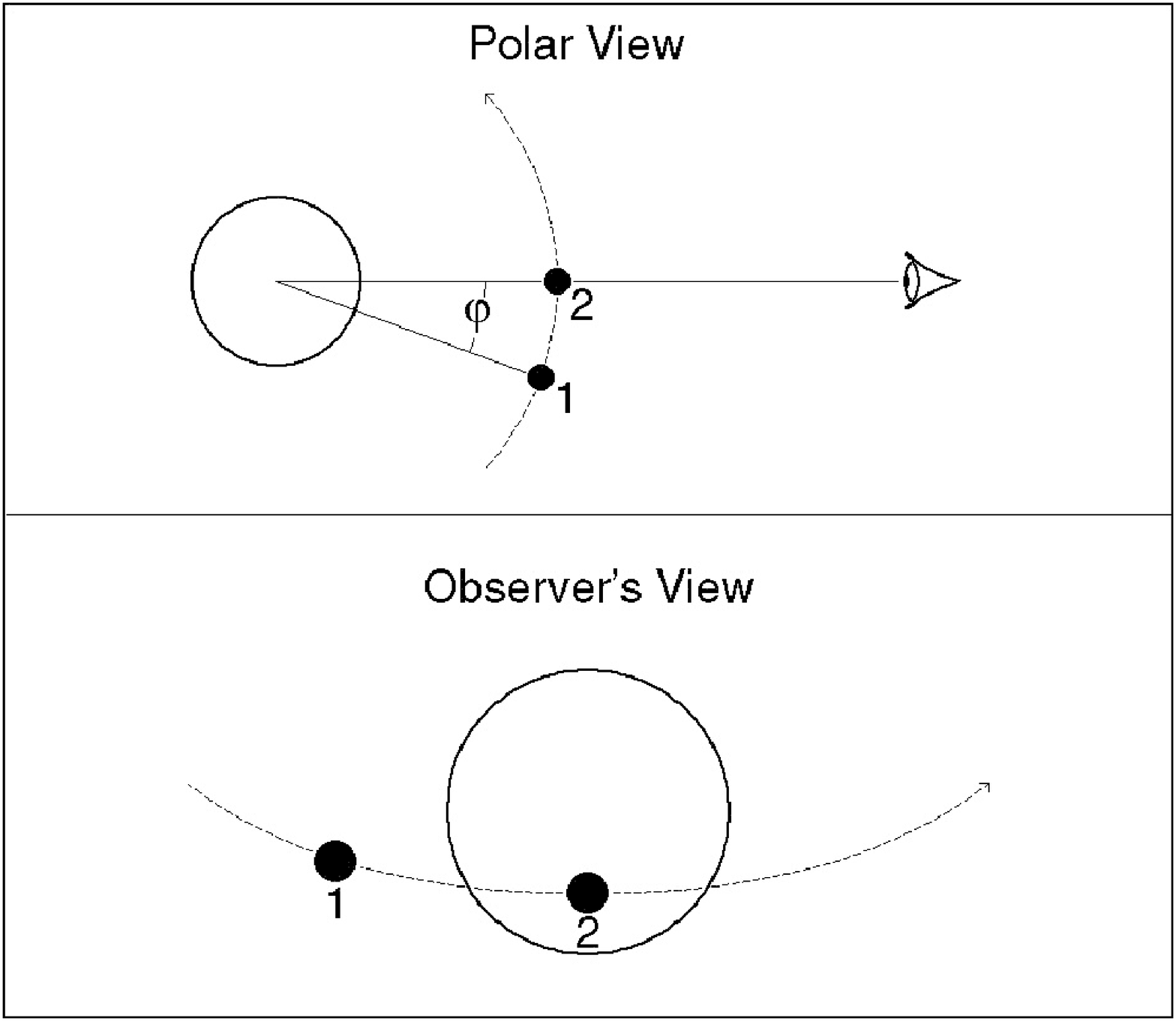}
\end{figure}

\begin{figure}
\plotone{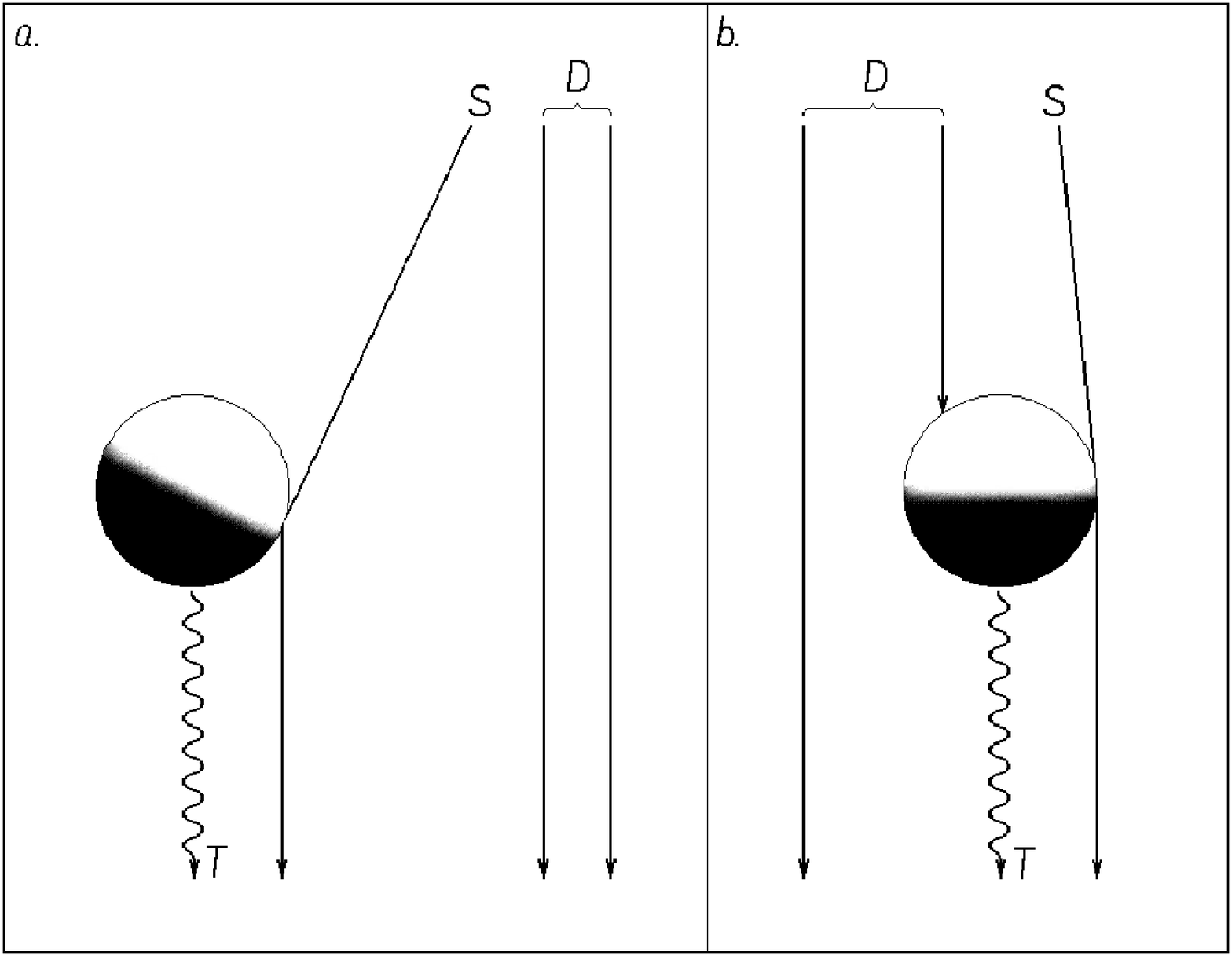}
\end{figure}

\begin{figure}
\plotone{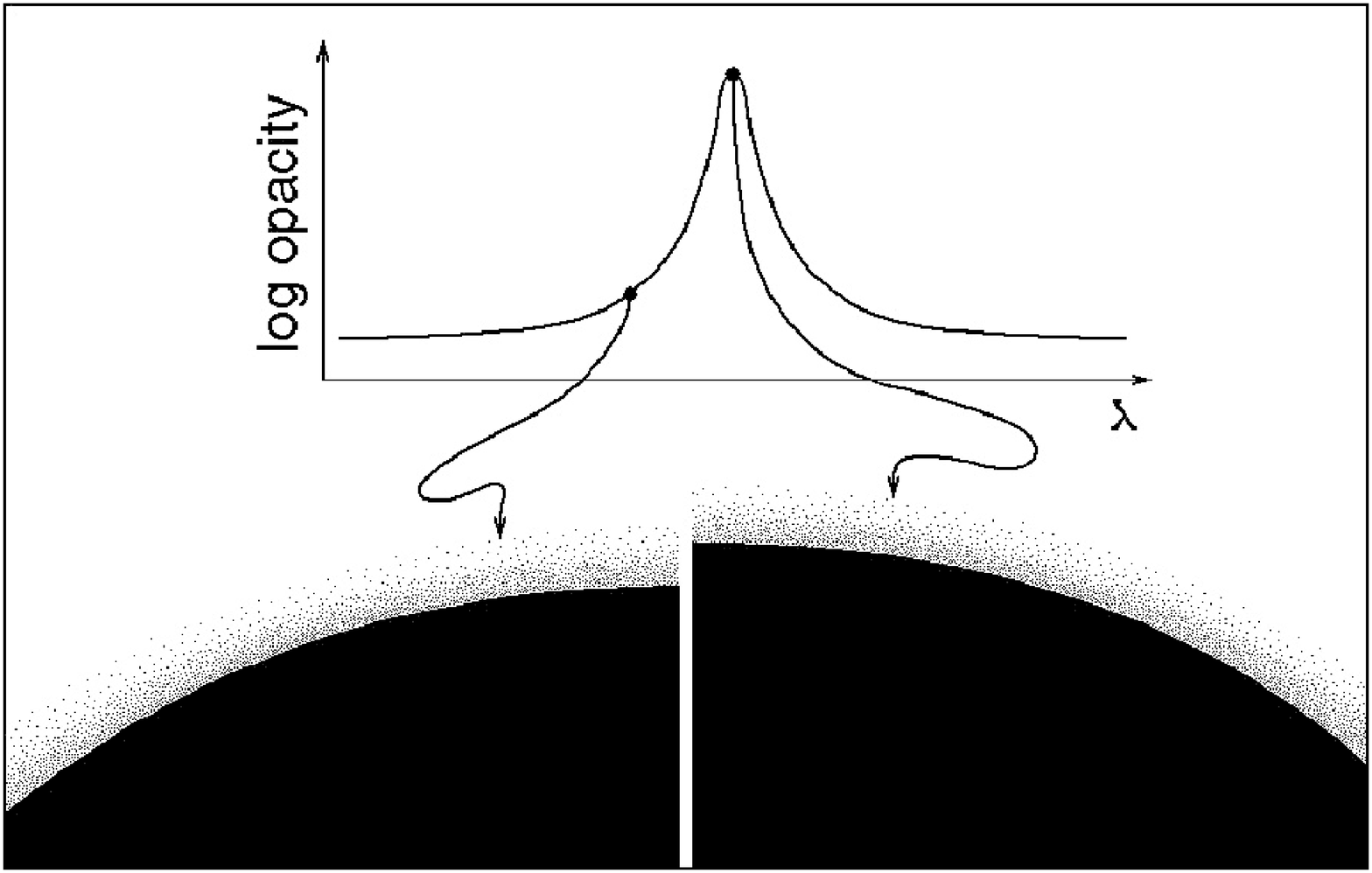}
\end{figure}

\clearpage

\begin{figure}
\plotone{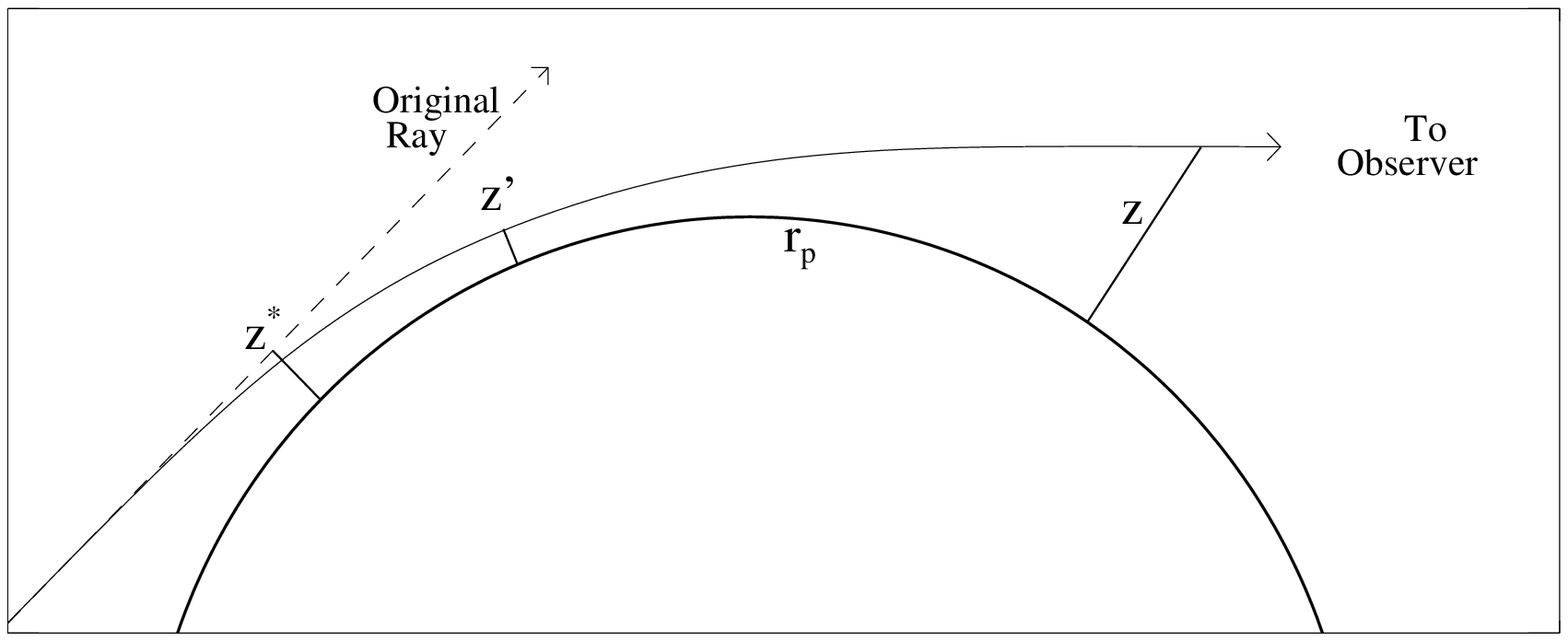}
\end{figure}

\clearpage

\begin{figure}
\plotone{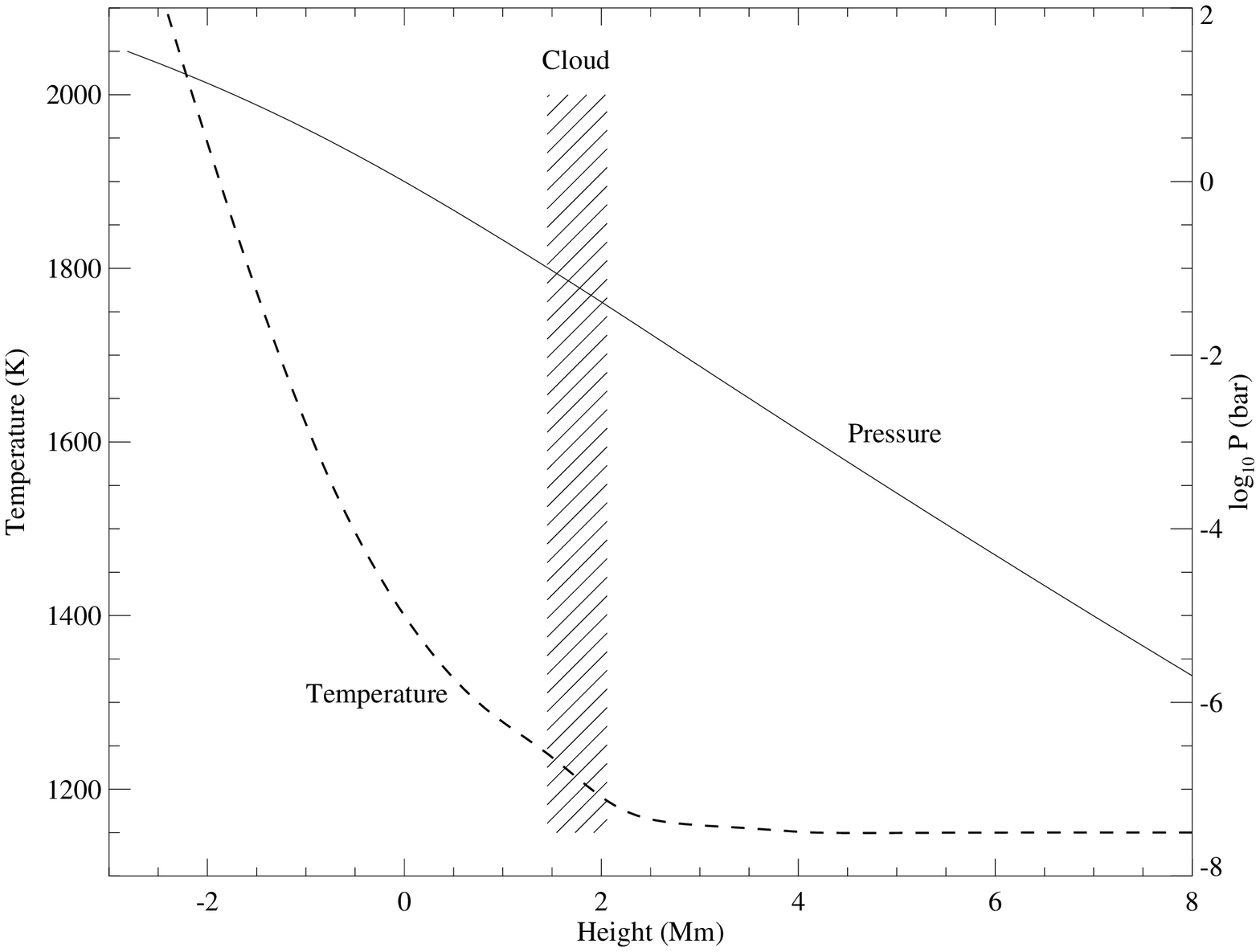}
\end{figure}

\begin{figure}
\plotone{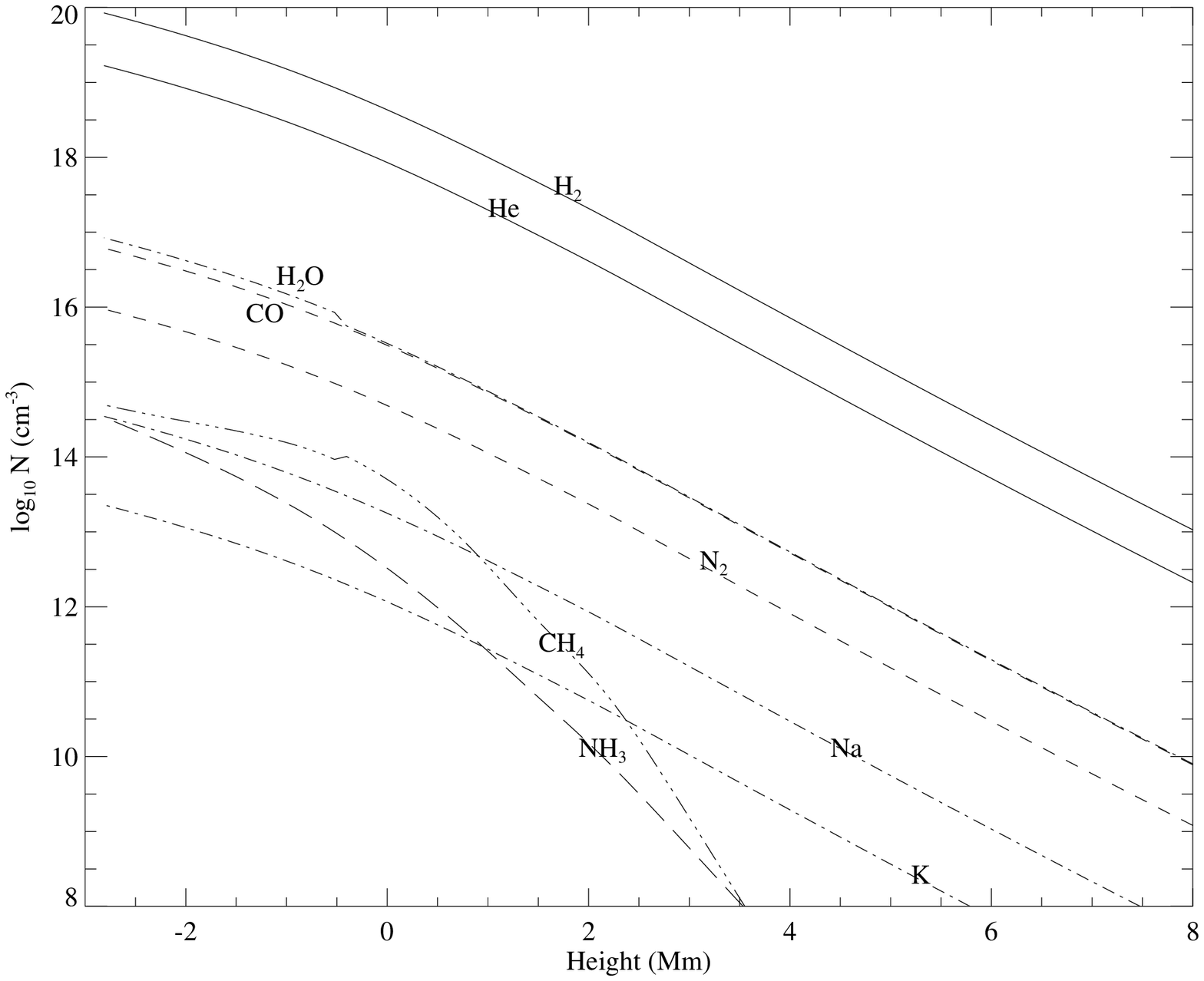}
\end{figure}

\begin{figure}
\plotone{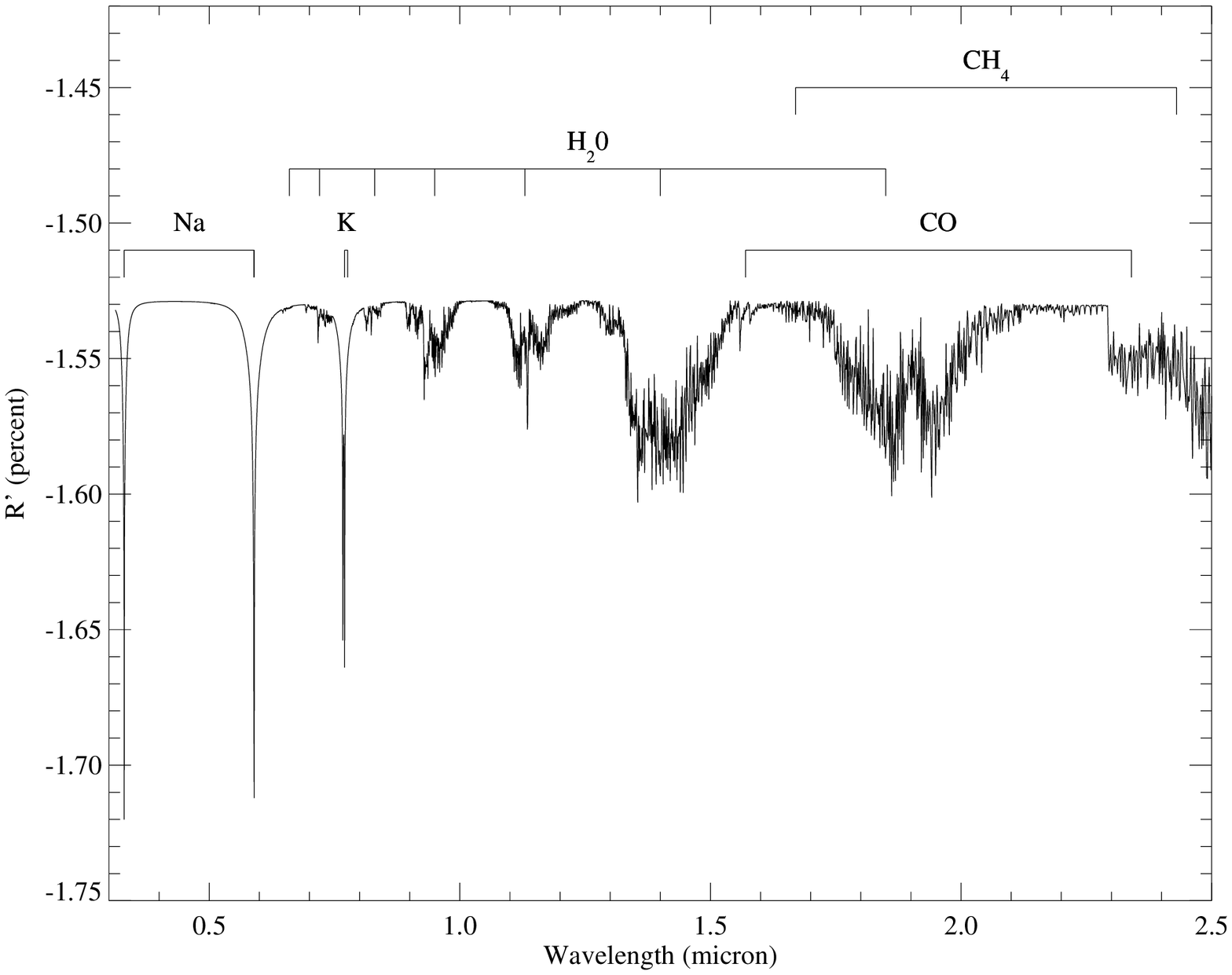}
\end{figure}

\begin{figure}
\plotone{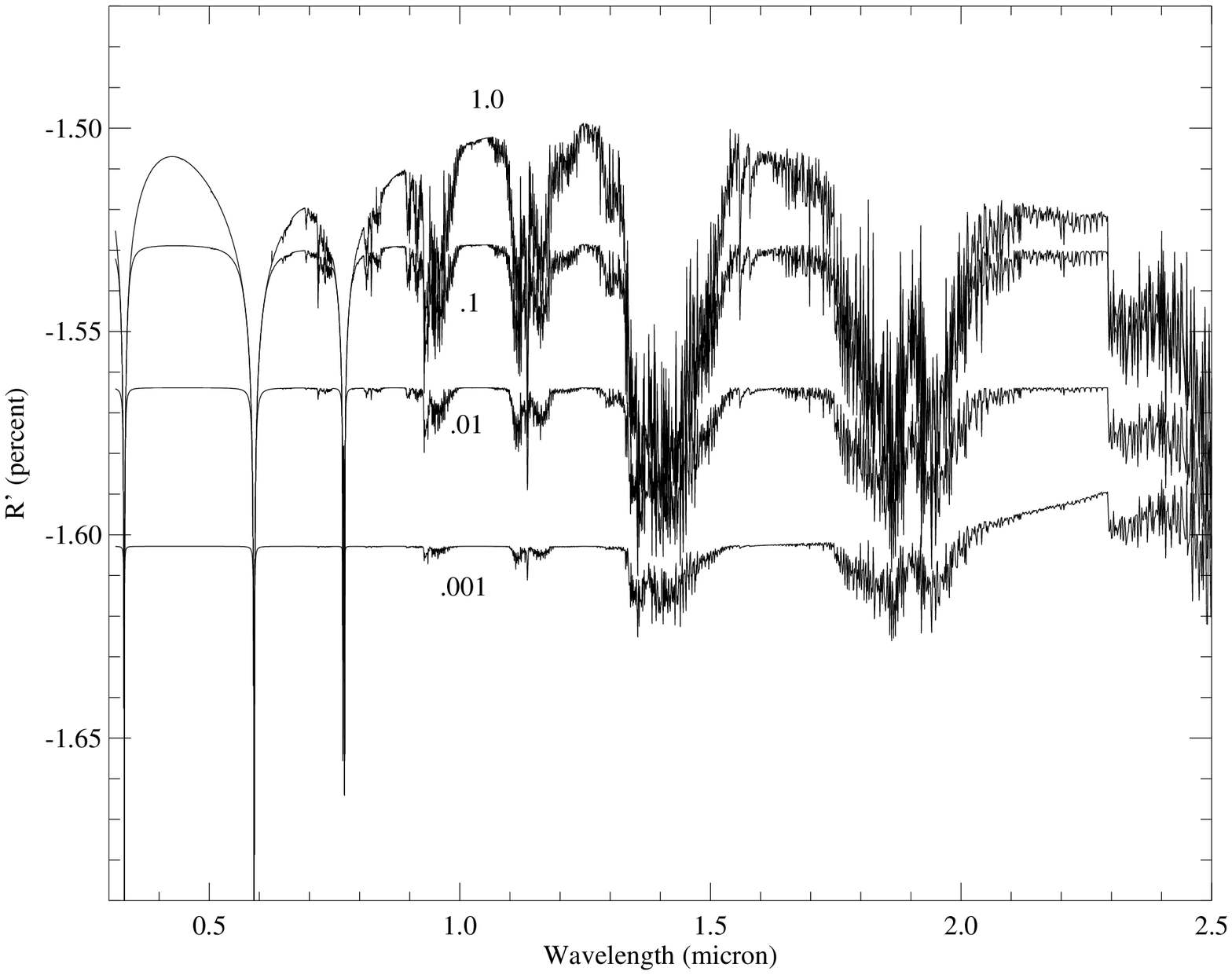}
\end{figure}

\begin{figure}
\plotone{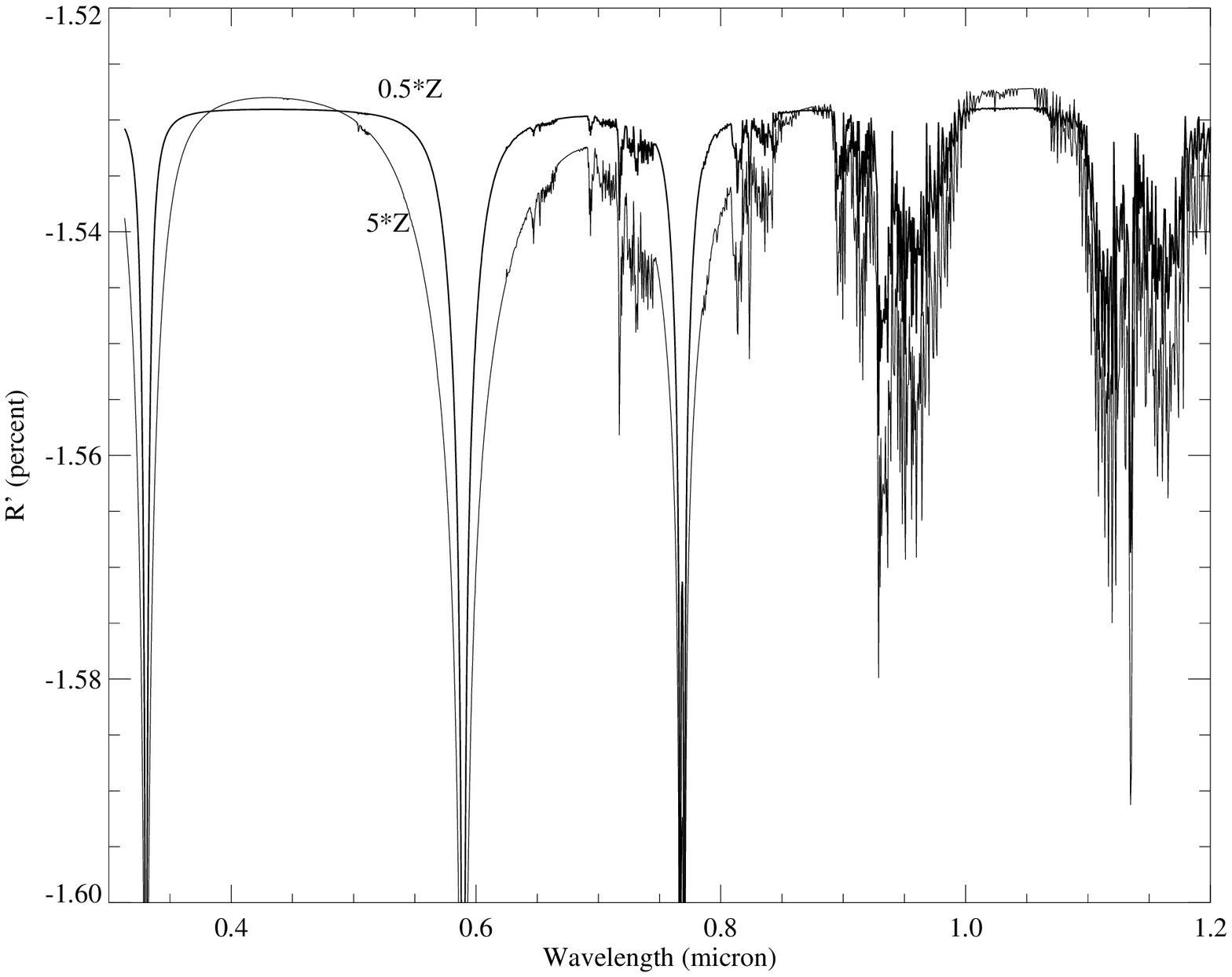}
\end{figure}

\begin{figure}
\plotone{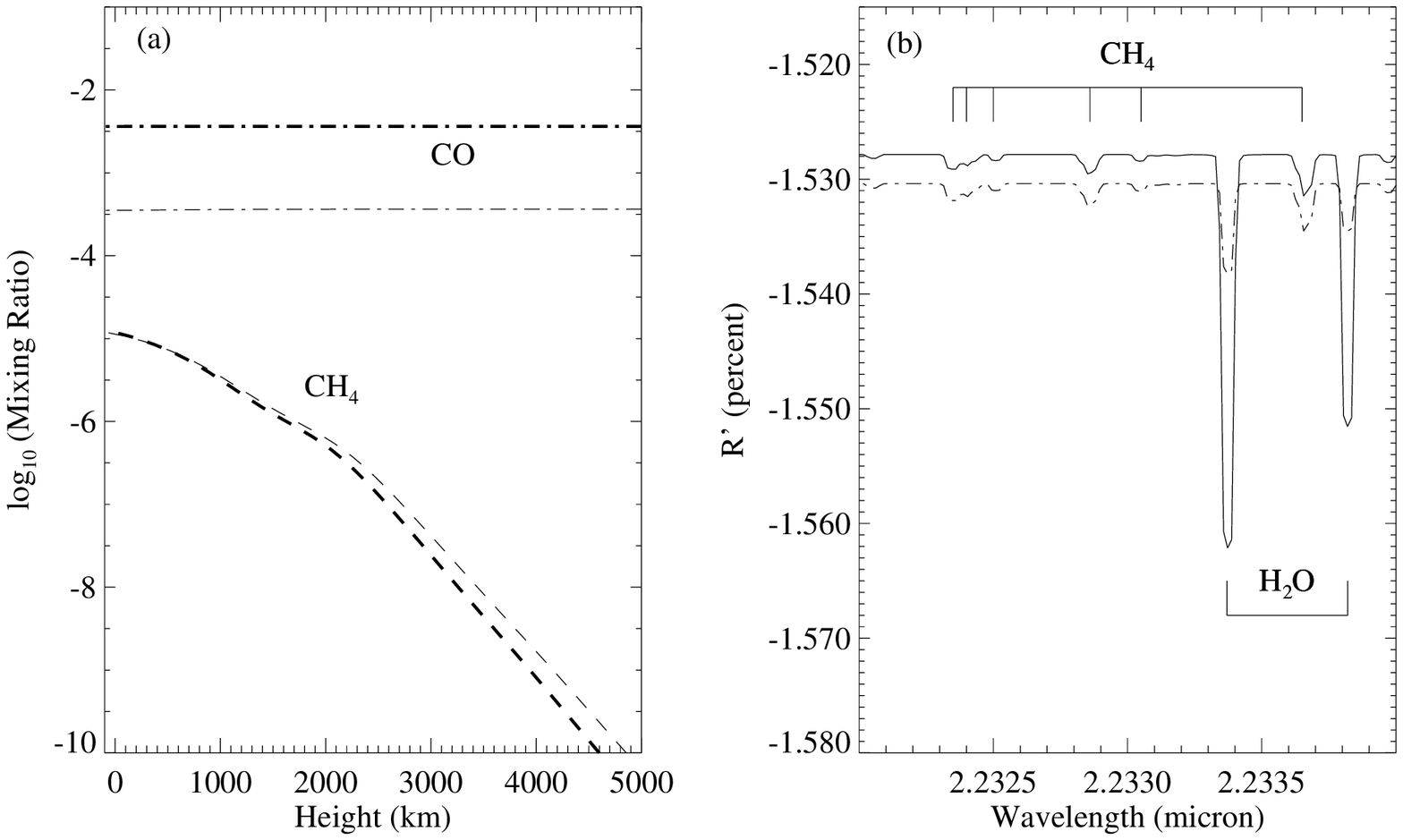}
\end{figure}

\begin{figure}
\plotone{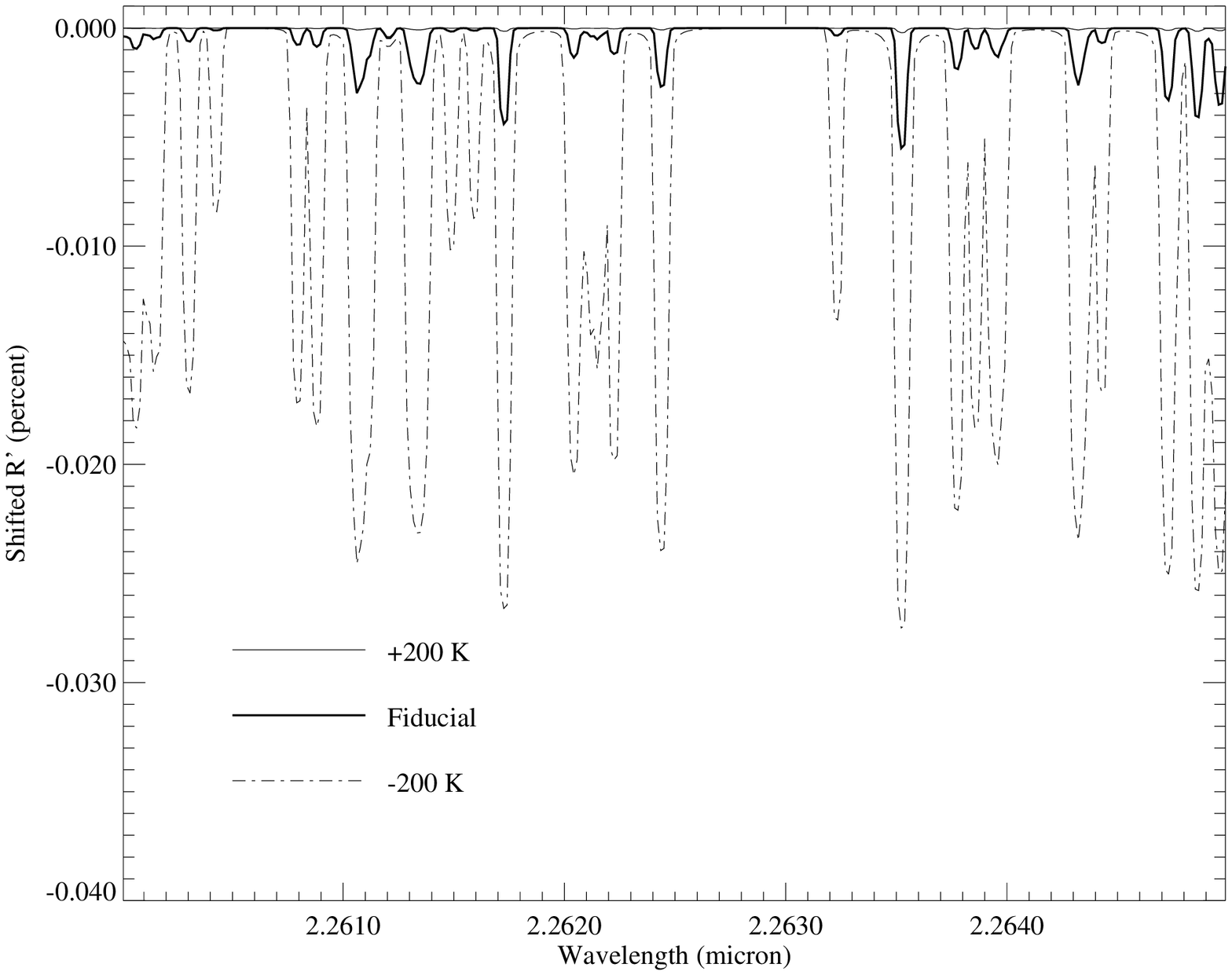}
\end{figure}

\begin{figure}
\plotone{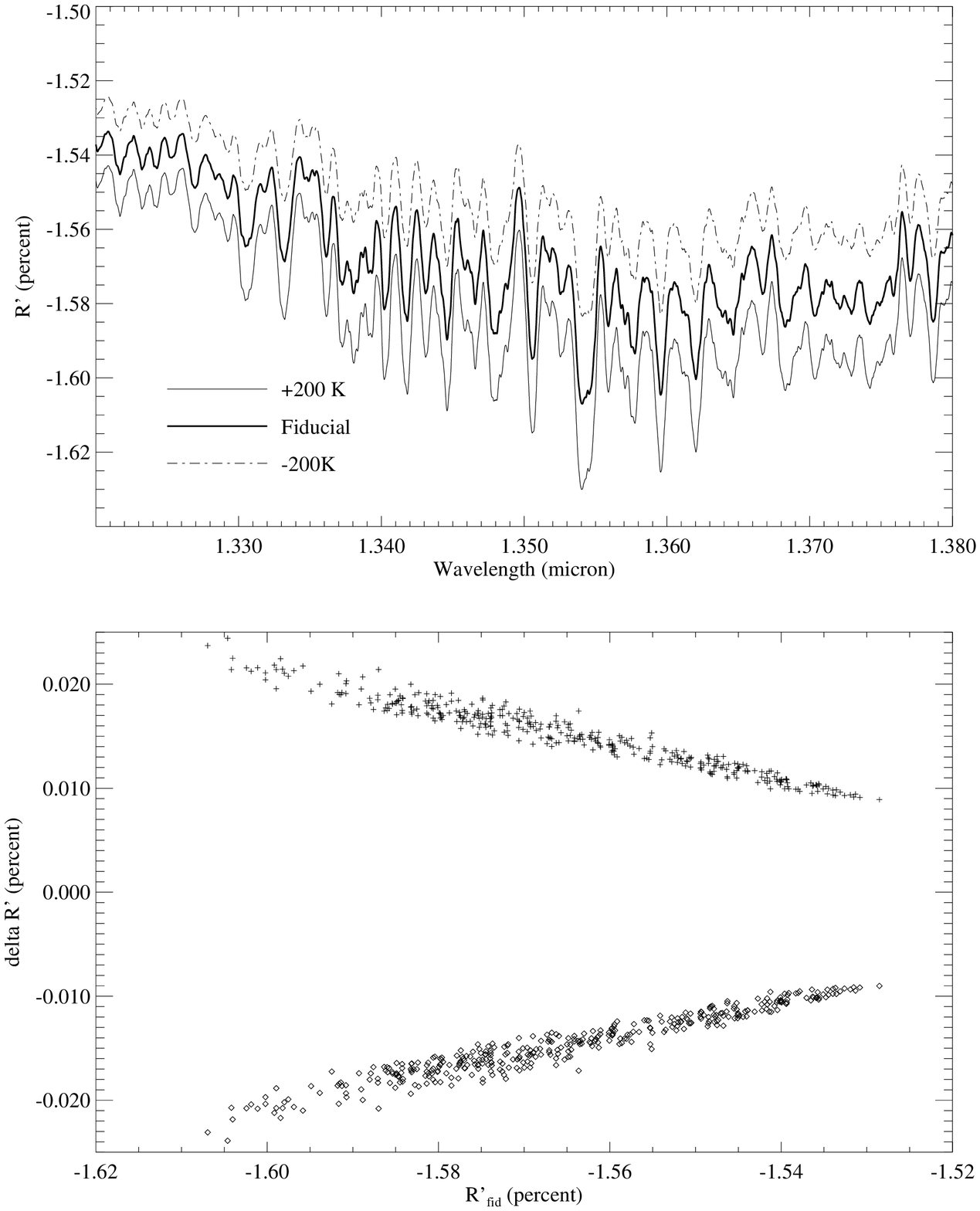}
\end{figure}

\begin{figure}
\plotone{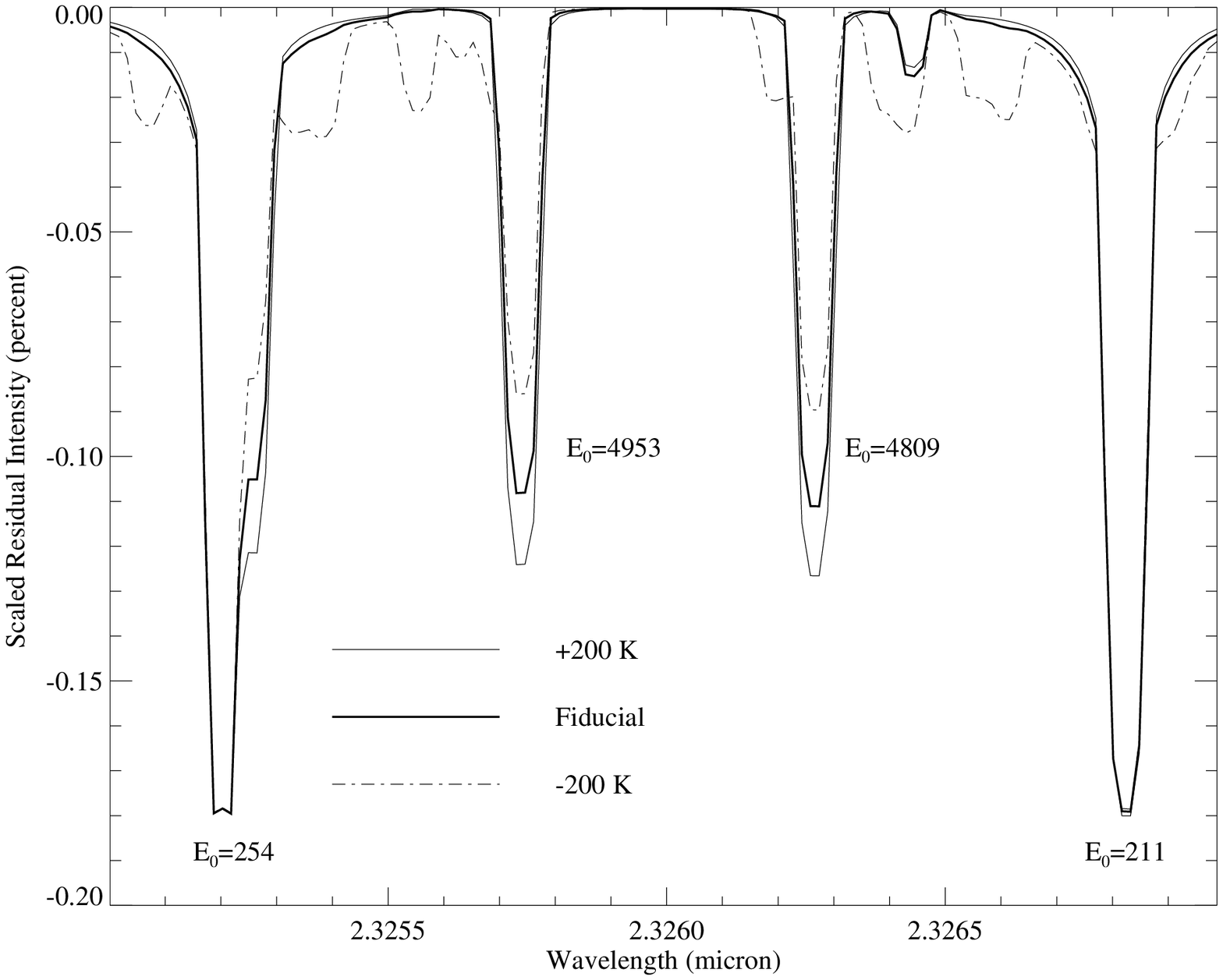}
\end{figure}

\begin{figure}
\plotone{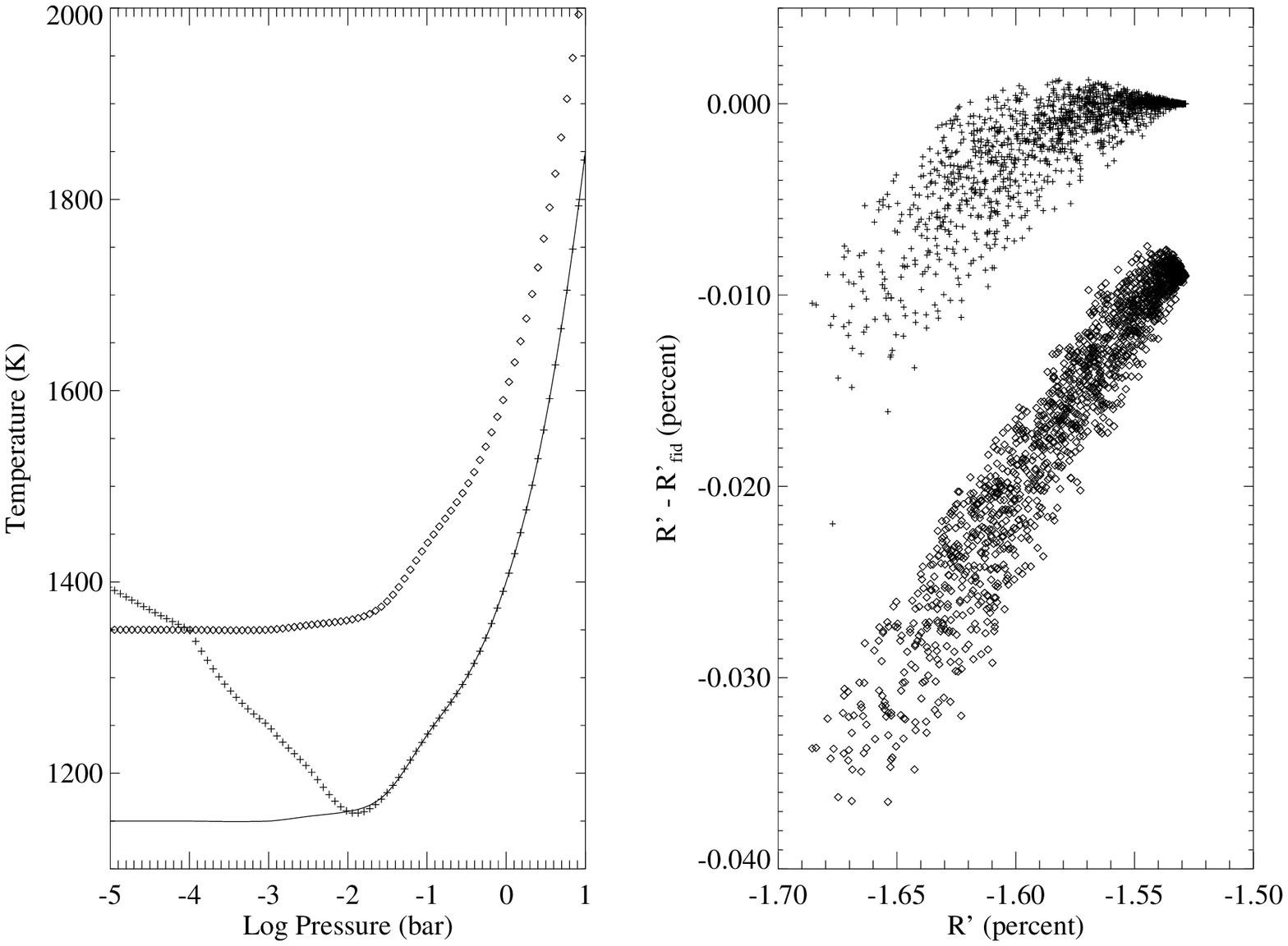}
\end{figure}

\begin{figure}
\plotone{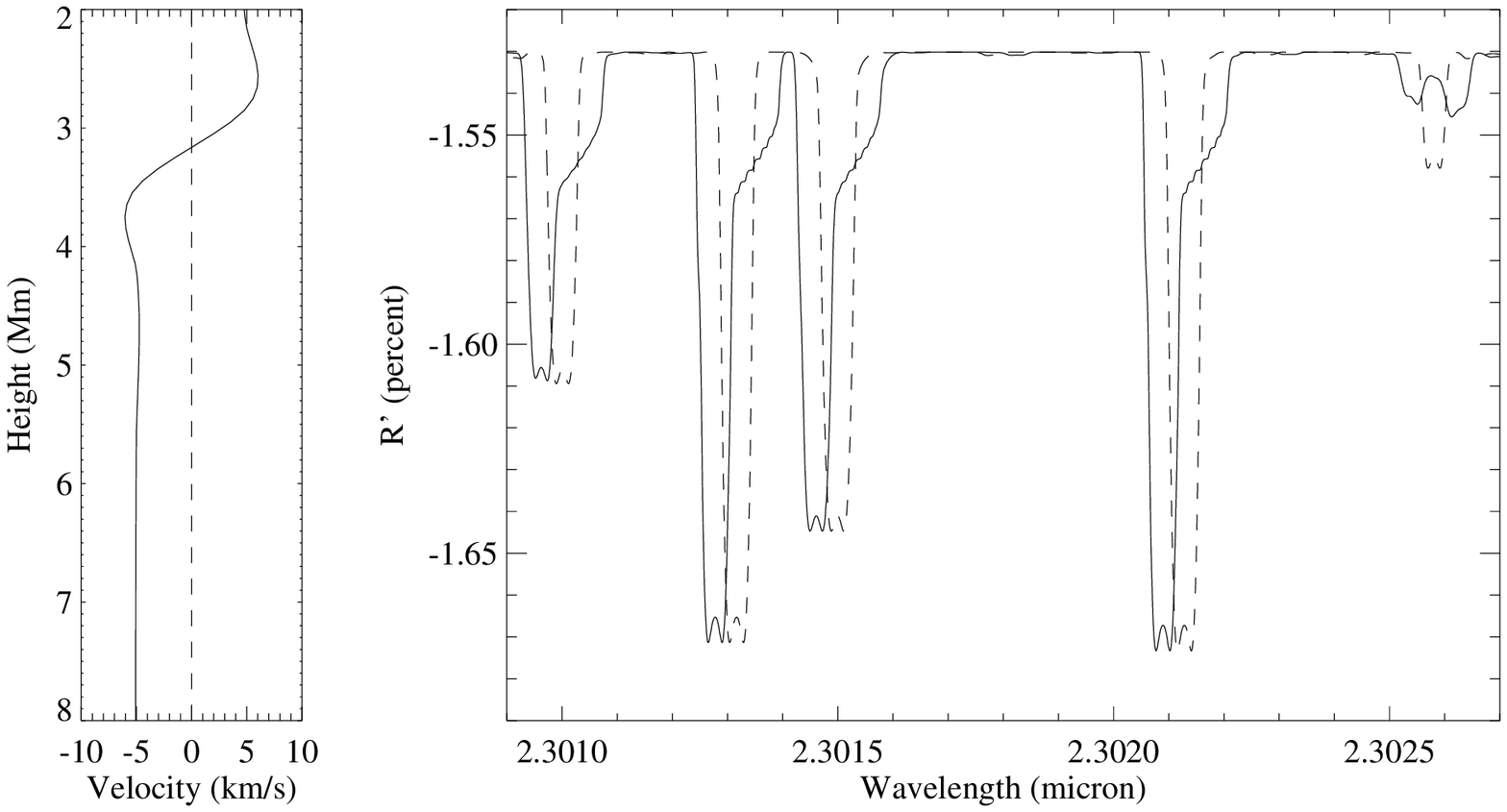}
\end{figure}

\clearpage
\begin{figure}
\plotone{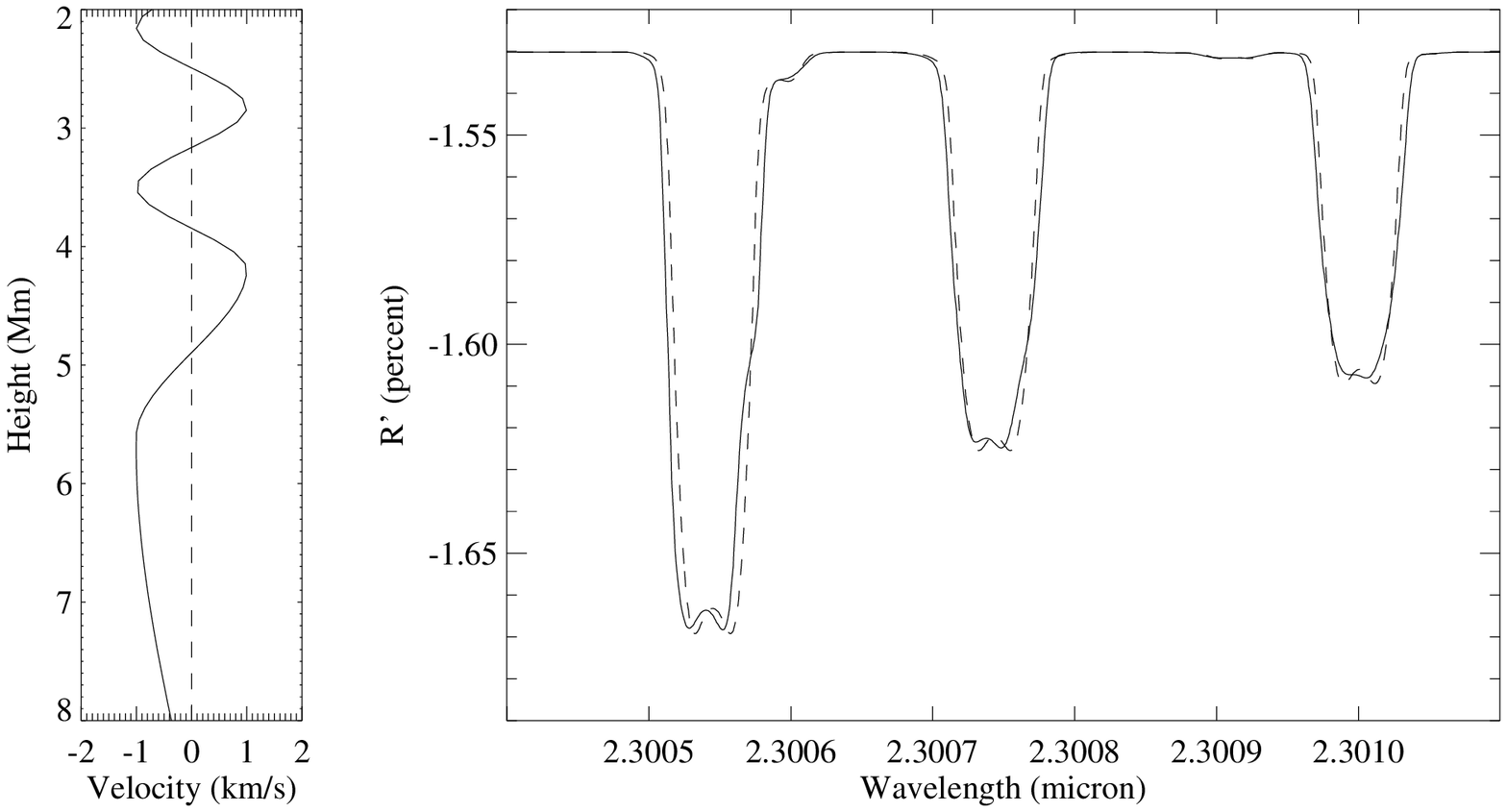}
\end{figure}

\clearpage
\begin{figure}
\plotone{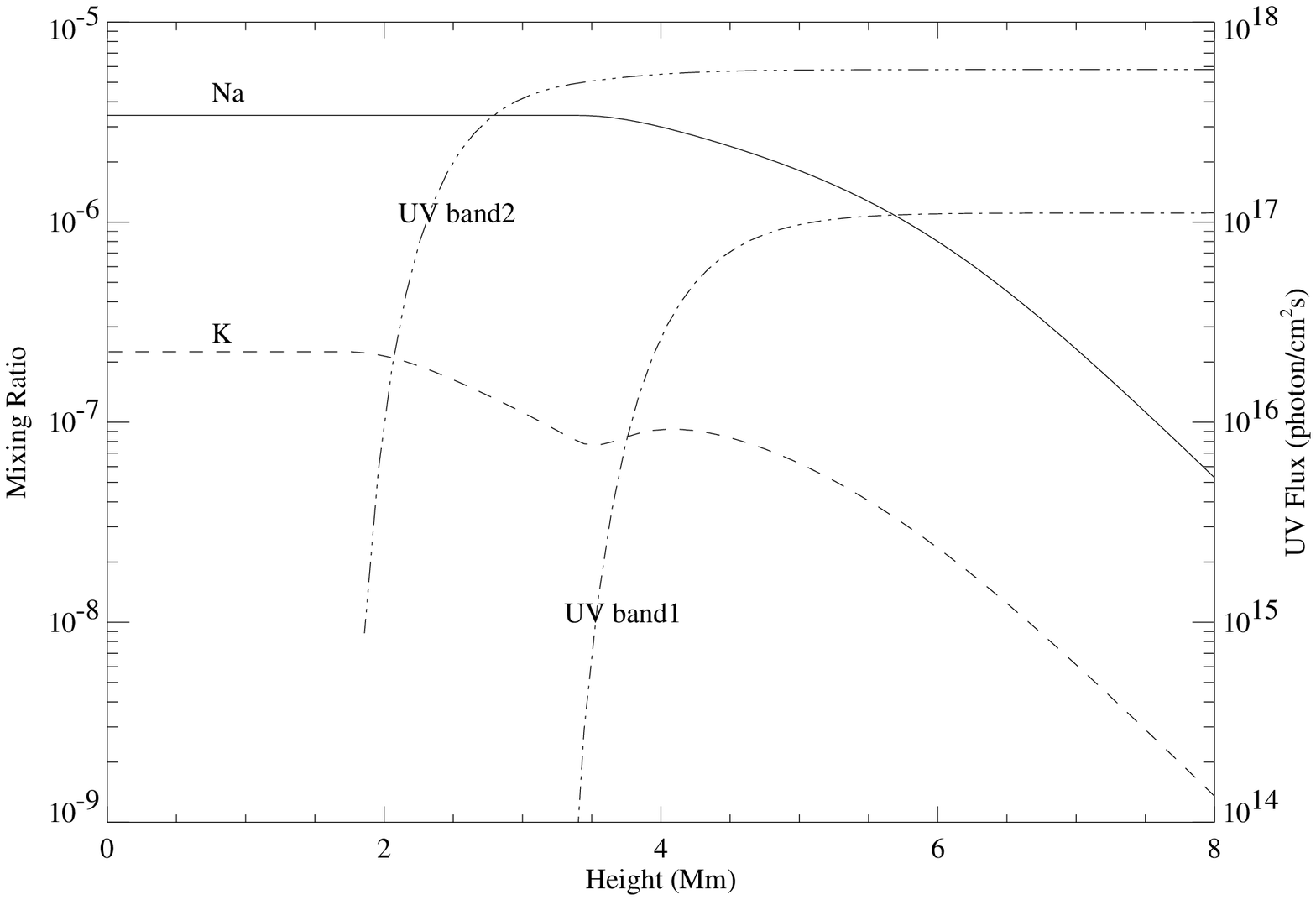}
\end{figure}

\begin{figure}
\plotone{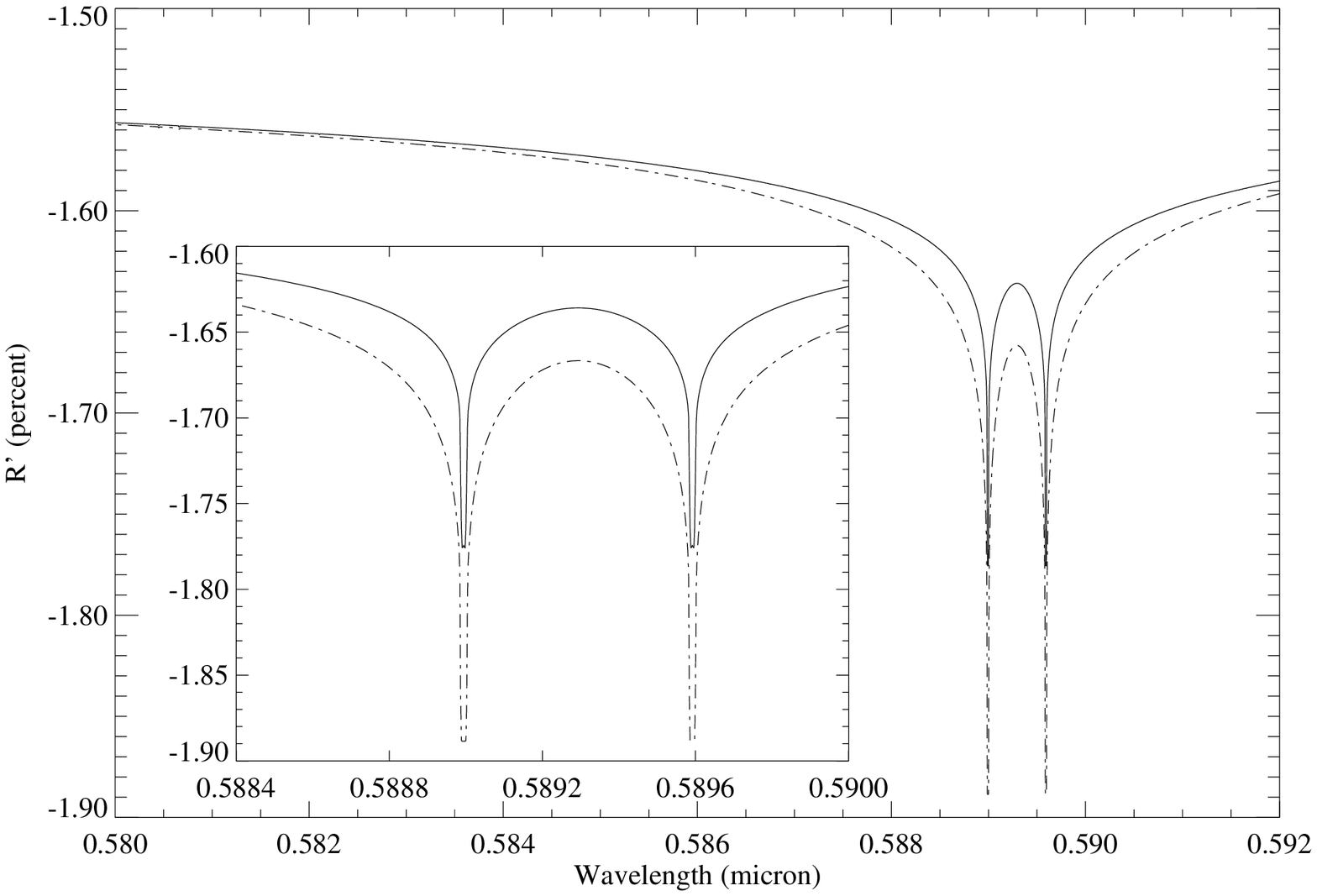}
\end{figure}

\begin{figure}
\plotone{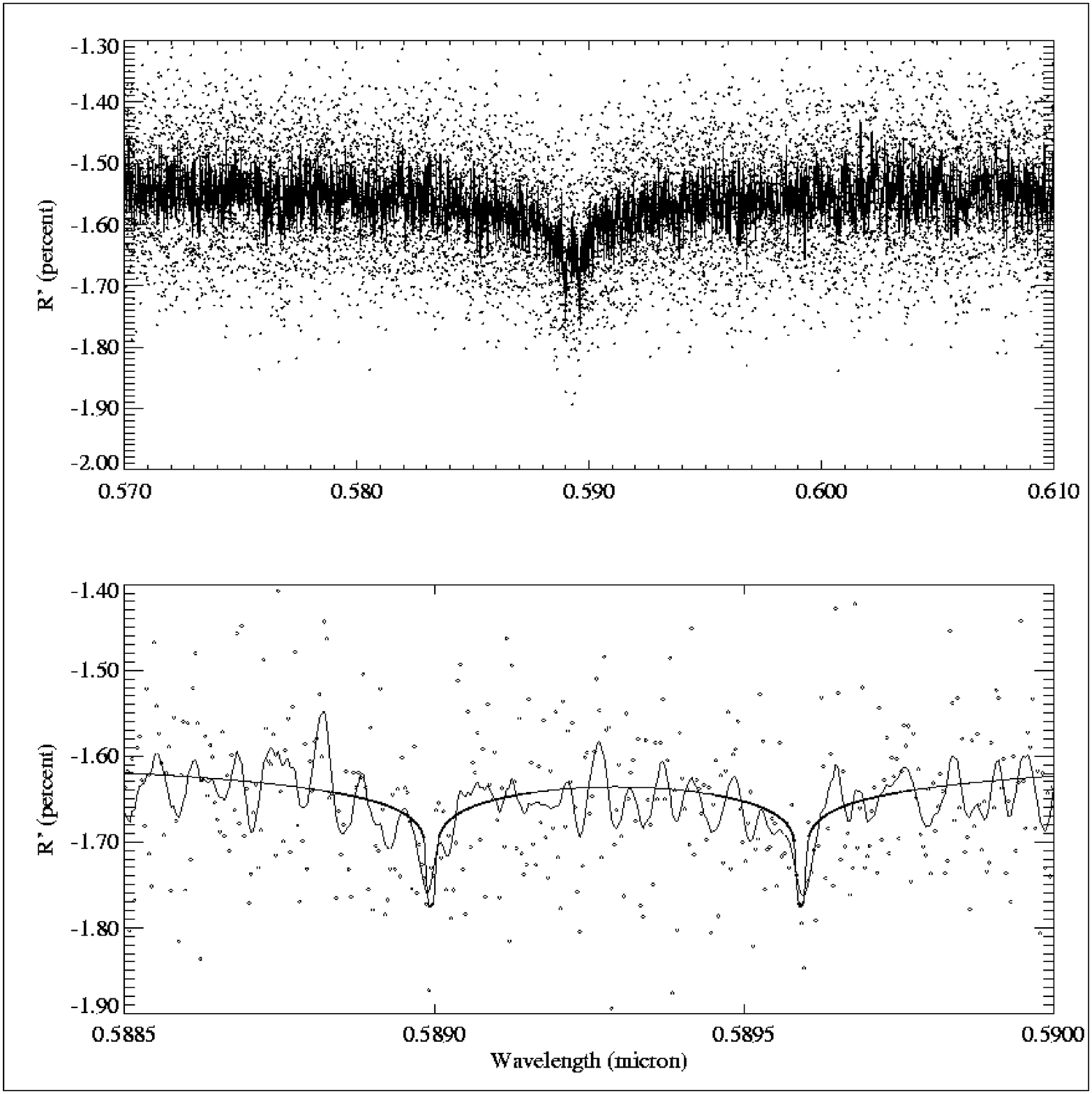}
\end{figure}

\begin{figure}
\plotone{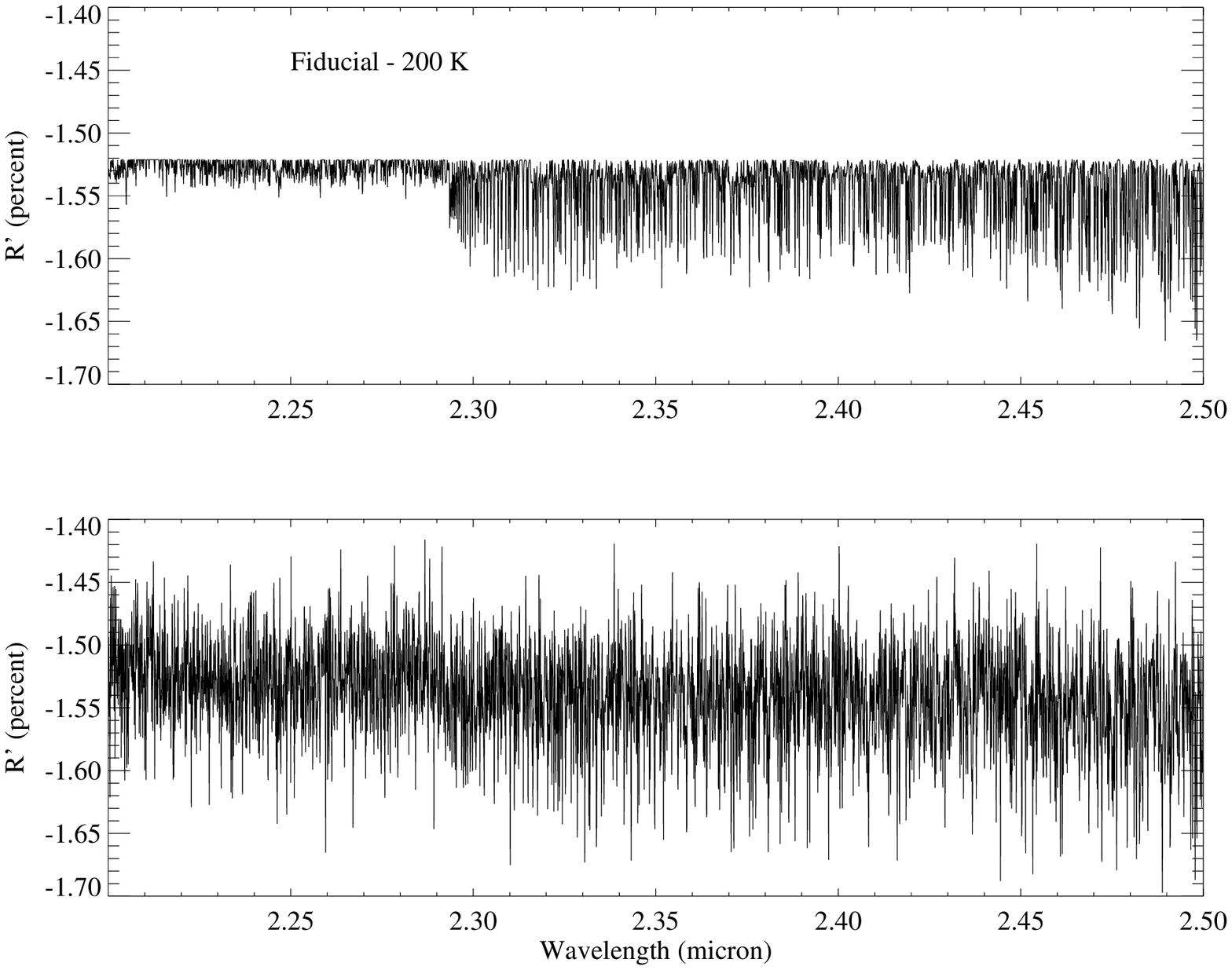}
\end{figure}

\begin{figure}
\plotone{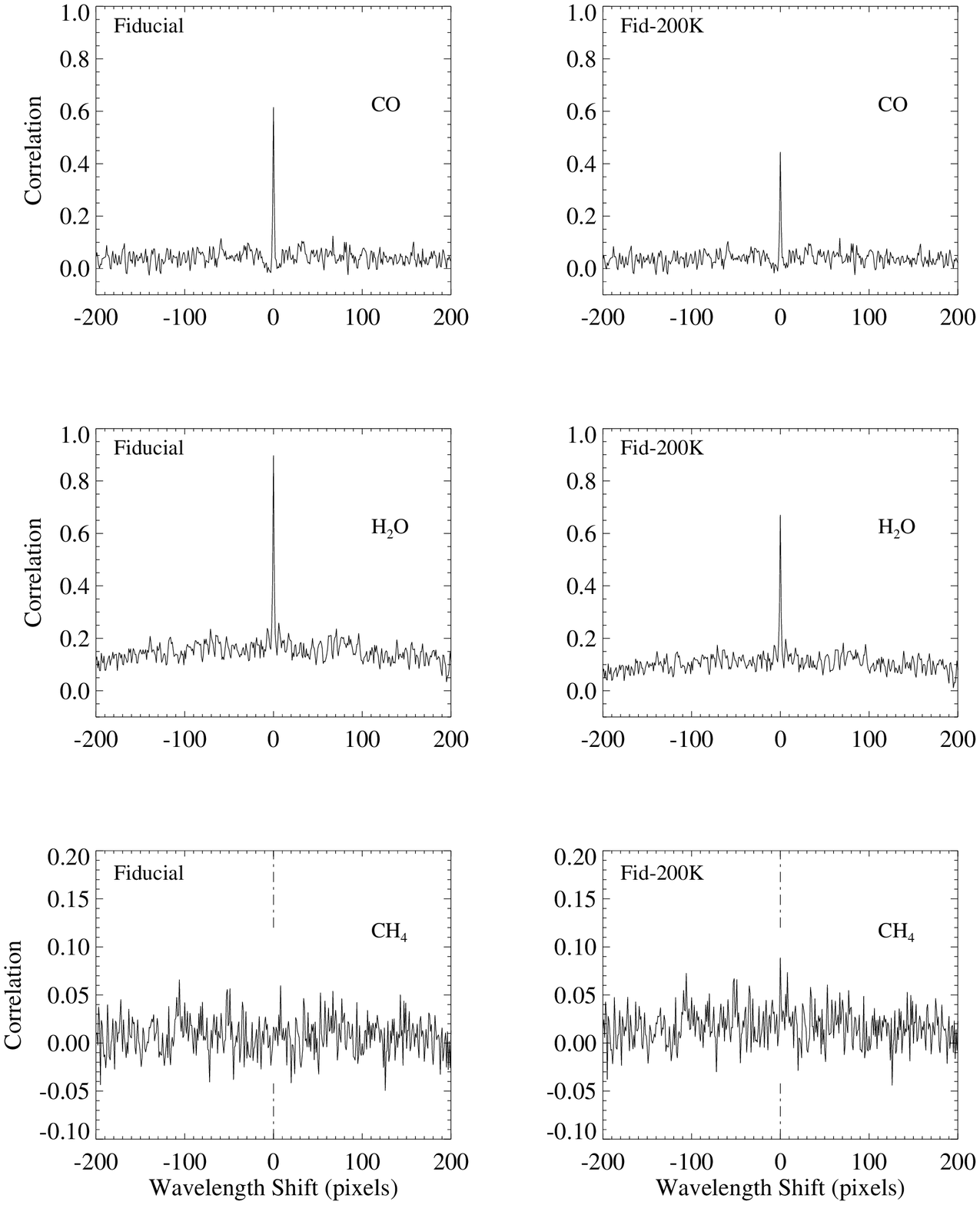}
\end{figure}

\begin{figure}
\plotone{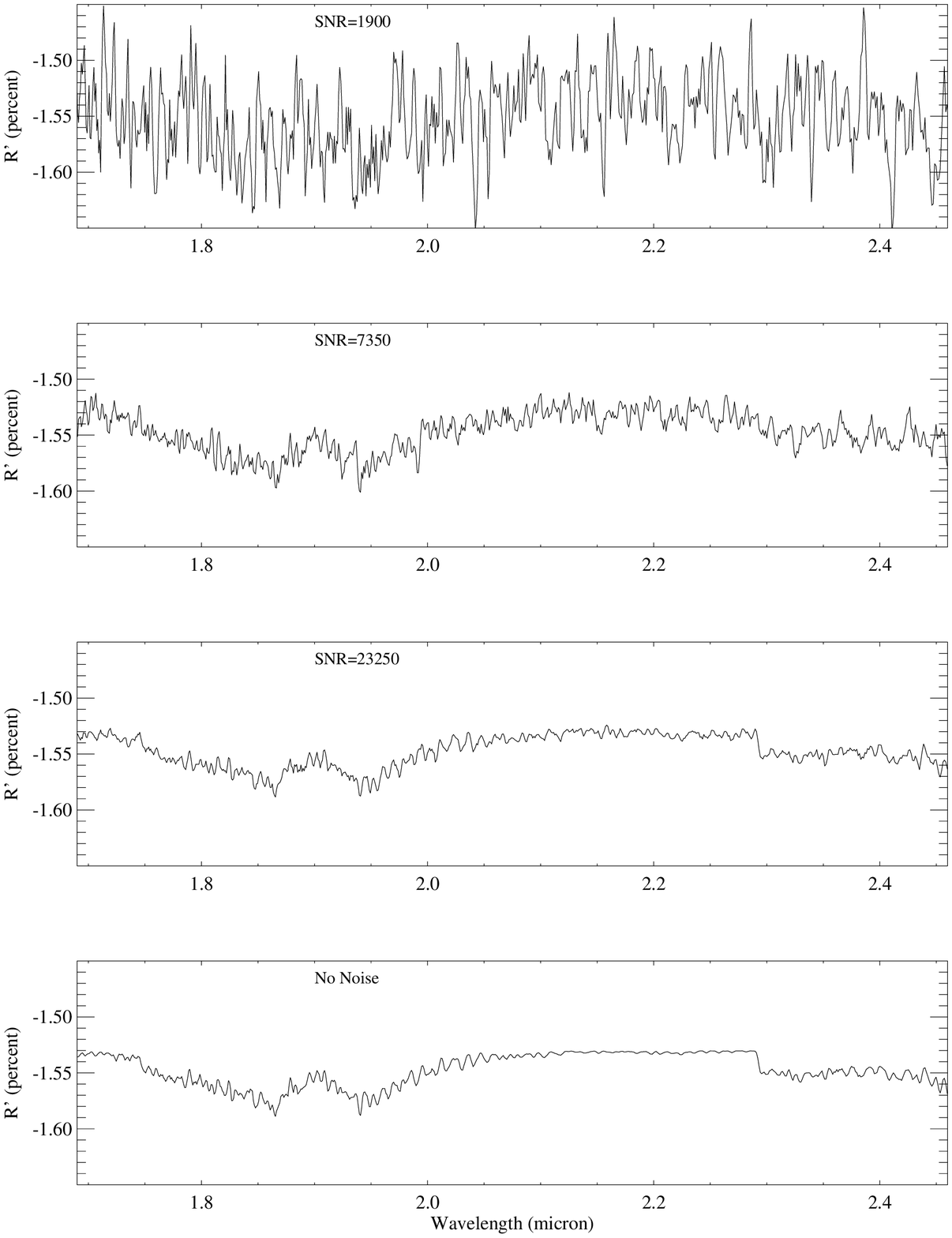}
\end{figure}

\clearpage
\end{document}